\documentclass[acmsmall,screen,nonacm]{acmart}

\definecolor{codegreen}{rgb}{0,0.6,0}
\definecolor{codegray}{rgb}{0.5,0.5,0.5}
\definecolor{codepurple}{rgb}{0.58,0,0.82}
\definecolor{backcolour}{rgb}{0.95,0.95,0.92}

\AtBeginDocument{%
  \providecommand\BibTeX{{%
    \normalfont B\kern-0.5em{\scshape i\kern-0.25em b}\kern-0.8em\TeX}}}

\usepackage[printonlyused]{acronym}

\newacro{mAPN}  [mAPN]  {multi-Alternative Process Network}
\newacro{KPN} [KPN]  {Kahn Process Network}
\newacro{SDF}   [SDF]   {Synchronous Dataflow}
\newacro{ASA} [ASA]  {Automatic Subtitling Application}
\newacro{MoC} [MoC]  {Model of Computation}
\newacro{ML} [ML]  {Machine Learning}
\newacro{UAV} [UAV]  {Unmanned Aerial Vehicle}
\newacro{AVP} [AVP]  {Autonomous Valet Parking}
\newacro{DLP} [DLP]  {Data Level Parallelism}
\newacro{TLP} [TLP]  {Task Level Parallelism}
\newacro{EPN} [EPN]  {Expandable Process Networks}

\newacro{ASP} [ASP] {Algorithm Selection Problem}

 \usepackage{xcolor}
\usepackage{ifthen}
\usepackage{tikz}
\usepackage{pifont}%
\usepackage{subcaption}

\usepackage{physunits}

\definecolor{turkis}{rgb}{0.0, 0.55, 0.55}
\definecolor{darkcyan}{rgb}{0.0, 0.55, 0.55}
\definecolor{darkred}{rgb}{0.55, 0.0, 0.0}
\definecolor{darkpowderblue}{rgb}{0.0, 0.2, 0.6}
\definecolor{forestgreen(traditional)}{rgb}{0.0, 0.27, 0.13}
\definecolor{green(html/cssgreen)}{rgb}{0.0, 0.5, 0.0}
\definecolor{safetyorange(blazeorange)}{rgb}{1.0, 0.4, 0.0}
\definecolor{purplew}{rgb}{0.86, 0.79, 0.28}
\definecolor{darkmagenta}{rgb}{0.55, 0.0, 0.55}

\definecolor{forestgreenp}{rgb}{0.13, 0.55, 0.13}
\definecolor{internationalkleinblue}{rgb}{0.0, 0.18, 0.65}

\newcommand{\greenline}{\raisebox{1pt}{\tikz{\draw[-,forestgreenp,solid,line width = 3pt](0,0 ) -- (2mm,0);}}}

\newcommand{\turkisline}{\raisebox{1pt}{\tikz{\draw[-,turkis,solid,line width = 3pt](0,0 ) -- (2mm,0);}}}
\newcommand{\blackline}{\raisebox{1pt}{\tikz{\draw[-,black,solid,line width = 3pt](0,0 ) -- (2mm,0);}}}
\newcommand{\blueline}{\raisebox{1pt}{\tikz{\draw[-,internationalkleinblue,solid,line width = 3pt](0,0 ) -- (2mm,0);}}}
\newcommand{\purpline}{\raisebox{1pt}{\tikz{\draw[-,darkmagenta,solid,line width = 3pt](0,0 ) -- (2mm,0);}}}

\newcommand{\orangeline}{\raisebox{1pt}{\tikz{\draw[-,orange,solid,line width = 3pt](0,0 ) -- (2mm,0);}}}

\newtheorem{definition}{Definition}[section]

\usepackage{algorithm2e}
\RestyleAlgo{ruled}

\SetCommentSty{mycommfont}

\begin{document}

\title{mAPN: Modeling, Analysis, and Exploration of Algorithmic and Parallelism Adaptivity}

\author{Hasna Bouraoui}
\email{hasna.bouraoui@tu-dresden.de}
\orcid{0000-0002-2832-1979}
\affiliation{%
  \institution{Technische Universit\"{a}t Dresden}
  \country{Germany}
}

\author{Chadlia Jerad}
\email{chadlia.jerad@ensi-uma.tn}
\orcid{https://orcid.org/0000-0002-5442-3098}
\affiliation{%
  \institution{University of Manouba}
  \streetaddress{ENSI, Campus Universitaire de la Manouba}
  \country{Tunisia}
}
\author{Omar Romdhani}
\email{omar.romdhani@ensi-uma.tn }
\orcid{https://orcid.org/0000-0002-7858-3870}
\affiliation{%
  \institution{University of Manouba}
  \streetaddress{ENSI, Campus Universitaire de la Manouba}
  \country{Tunisia}
}
\email{omar.romdhani@ensi-uma.tn}

\author{Jeronimo Castrillon}
\email{jeronimo.castrillon@tu-dresden.de}
\orcid{0000-0002-5007-445X}
\affiliation{%
  \institution{Technische Universit\"{a}t Dresden}
  \country{Germany}
}

\renewcommand{\shortauthors}{Bouraoui and Jerad, et al.}

\begin{abstract}

Using parallel embedded systems these days is increasing. They are getting more complex due to integrating multiple functionalities in one application or running numerous ones concurrently. This concerns a wide range of applications, including streaming applications, commonly used in embedded systems. These applications must implement adaptable and reliable algorithms to deliver the required performance under varying circumstances (e.g., running applications on the platform, input data, platform variety, etc.). 

Given the complexity of streaming applications, target systems, and adaptivity requirements, designing such systems with traditional programming models is daunting. This is why model-based strategies with appropriate \ac{MoC} have long been studied for embedded system design.
Dataflow models, in particular, are a good fit for streaming applications that execute in parallel on embedded devices. 
Most of today's models, however, are based on static dataflow models with adaptivity extensions to describe data parallelism. Some dynamic dataflows capture dynamic behavior but offer only limited support for algorithmic adaptivity. This work provides algorithmic adaptivity on top of parallelism for dynamic dataflow
to express larger sets of variants and trade-offs. We present a multi-Alternative Process Network (mAPN), a high-level abstract representation in which several variants of the same application coexist in the same
graph expressing different implementations. We introduce \ac{mAPN} properties and its formalism to describe various local implementation alternatives.  Furthermore, mAPNs are enriched with metadata to provide the alternatives with quantitative annotations in terms of a specific metric. 
To help the user analyze the rich space of variants, we propose a methodology to extract feasible variants under user and hardware constraints.
At the core of the methodology is an algorithm for computing global metrics of an execution (e.g., execution time) of different alternatives from a compact \ac{mAPN} specification.  We validate our approach by exploring several possible variants created for the \ac{ASA} on two hardware platforms and analyzing the trade-off space. A comparison to the well-known analysis tool $SDF^3$ is also performed, where we showed that we are more than 500 times faster for a large number of variants. 

\end{abstract}



\maketitle
\section{Introduction}
\label{sec:introduction}

The complexity of applications aiming to solve nowadays problems is naturally increasing across domains. Many sectors are concerned: from automotive applications, through medical and health systems, to wireless communication (e.g., 5G and beyond), and ending with applications requiring biometric authentication (e.g., speaker recognition, iris recognition, automatic subtitling).

For instance, highly automated systems are installed in today's vehicles in the automotive sector. These systems, commonly referred to as ADAS (Advanced Driving Assistance Systems)~\cite{eskandarian2012handbook, bergenhem2012overview, sahayadhas2012detecting}. The complexity of applications increases even further in Autonomous Driving (AD) scenarios. Several functionalities are needed in such applications, including capturing a scene, detecting, and object tracking (e.g., traffic signs, pedestrians). To serve this purpose, different technologies are used: from advanced sensing (e.g., Radars, LiDARs, cameras), to advanced algorithms (e.g., computer vision techniques with hard time constraints, machine learning techniques), and advanced actuation executing commands on time and precisely~\cite{yurtsever2020survey}.

Another example %
of a growing class of applications are those based on the biometric recognition. 
In automatic subtitling for video, for instance, functionalities like speaker recognition, speaker diarization, and speech recognition are important biometric algorithms. 
In the case of a live broadcast, automatic subtitling has to execute under real time constraints. 
In the aforementioned examples, a complex application is composed of several algorithms,
each exposing different trade-offs, making the application graph large in size and with complex architectures. %
This gain in complexity makes it hard from a developer's point of view to choose the proper implementation for her application beforehand. 
This becomes even harder considering the diversity of possible target hardware architectures, as the achieved implementation performance may significantly differ from one target to the other. 
The well known \ac{ASP}~\cite{rice1976algorithm} addressed this issue by combining the solution from existing algorithms rather than developing a new specific implementation for some problems. The idea is to identify the most adequate algorithm for a given problem under changing circumstances. However, the solution space is ample and makes depicting the suitable algorithm task daunting.  
It is challenging to manually select an adequate algorithm under user and hardware constraints in any given situation characterized by available hardware resources and energy budget.
To design such complex applications, embedded programmers need to understand algorithmic variants implementing the same functionality (i.e. algorithmic adaptivity) and how can they be deployed in parallel into possibly different manycore platforms (i.e. parallelism adaptivity). 
These applications benefit from enabled pipelining and task parallelism to execute on embedded manycore platforms (homogeneous or heterogeneous) using appropriate \ac{MoC}~\cite{ptolemyBook}. %

Dataflow is a Model of Computation (MoC) that inherently describes exploitable, yet explicit software parallelism at a high level of abstraction. 
Together with the pervasiveness of multicore hardware platforms, these features make dataflow models well suited for the description and analysis of a wide variety of applications, notably signal processing and streaming applications. 
Different variants of this model have been developed and studied, enabling thus their use in various frameworks and tools, such as LabVIEW, Scade, and Simulink. 
Dataflow models such as Synchronous Dataflow (SDF~\cite{lee87sdf}) and Cyclo-static dataflow (CSDF~\cite{csdf96}) are called static and enjoy predictability, solid formal properties, and amenability to powerful optimization techniques.
Dynamic dataflow graphs (e.g. Parameterized dataflow Graphs~\cite{bhattacharya2001parameterized}, SADF~\cite{stuijk2011scenario}, Parametrized Interfaced DF ~\cite{desnos2014pisdf,madronal2019papify}), however, allow for the variation of the consumption and production rates of actors at runtime.
Consequently, they are not entirely predictable at compile time.
Kahn Process Networks are a prominent example of dataflow MoC that is more expressive than the aforementioned MoCs~\cite{stuijk2011scenario} while being deterministic~\cite{gilles1974semantics}.
Works such as in~\cite{khasanov2018implicit} and~\citep{schor2014adapnet} target dynamic behavior of KPN in the sense of implicit support of data level parallelism. In ~\cite{ansel2009petabricks}, adaptivity is taken into account at the compiler level, but their dynamic MoC suffers from a lack of intuitiveness and usability.

This paper presents a novel model, mAPN multi-Alternative Process Networks (mAPN~\cite{bouraoui_et_al:OASIcs.PARMA-DITAM.2021.1}). mAPN is a dynamic MoC that supports algorithmic and parallelism adaptivity. It captures multiple algorithmic implementation variants, beyond what existing models allow, in a compact single-source specification. Thus allowing to express more than one possible implementation of steaming applications in the same graph. Such a model helps the user to study the performance of potential implementations at a high level of abstraction. This paper aims to show how can mAPN help the user exploring such complex design space, while expressing algorithmic and parallelism adaptivity, to achieve desired performance and meet hardware constraints. We also position mAPN and its tooling against state-of-the-art works on the topic of adaptivity. 
Concretely, our contributions are: 
\begin{itemize}
	\item We give a thorough analysis of 2 complex real world applications (c.f. section~\ref{sec:motivationalExamples}), which are the Automatic Subtitling Application (ASA) and Advanced driving-assistance systems (ADAS).
		 
	\item We provide a revised mAPN formalism (c.f. section~\ref{sec:formalism}).
	The extended formalism enables faster exploration, and includes support for a compact and implicit representation of data level parallelism.
	The model is enriched with annotations that represent quantitative metrics of interest to the designer (e.g., execution time), in addition to annotations to control the amount of \ac{DLP}.
	
	\item We present a methodology for automatic exploration of algorithmic variants based on mAPN annotations and the constraints introduced by the designer (c.f. section~\ref{sec:methodology}). We present mAPN$^{TS}$ (mAPN tool suite), a tool that allows researchers and designers to explore a given mAPN graph and select the number of appropriate variants.
	
	\item We demonstrate the methodology for the ASA (c.f. section~\ref{sec:caseStudyASA}) by evaluating 224 alternatives represented in one graph. The selection of suited variants is based on annotations and respecting constraints. In addition, we analyze the fidelity of our model-based approach for the ASA use case. The assessment of the performance of mAPN$^{TS}$ against $SDF^3$ tool and SADF formalism is presented as well. For a large number of variants, we show that we reach more than 500 speedup in exploring the graph using mAPN$^{TS}$ compared to $SDF^3$. 
	
	\item Finally, we compare mAPN against other MoC  and state similarities and differences between them (c.f. section~\ref{sec:relatedWork}).
\end{itemize}

\section{Motivational Examples}
\label{sec:motivationalExamples}

Adapting to changing execution contexts and hardware constraints requires reasoning about different possible implementations of an application. Whenever a new algorithm has to be implemented for a specific hardware target, there is a general tendency nowadays to compose the solution from existing implementations rather than developing new algorithms. This is known as the \ac{ASP}~\cite{rice1976algorithm} and shifts the burden from finding the right solution to identifying the appropriate existing algorithms.
Defining the appropriate algorithm, however, is challenging since the solution space is exponentially ample. It depends on several changing circumstances, such as the user's desired performance and the available resources on the target platform. This is referred to as the no-free lunch theorem~\cite{wolpert1997no}, which states that no single algorithm has the best overall performance. 
Today, a common practice is to apply methods that automatically determine which strategy to employ in these adaptive applications scenarios. Many works in the literature addressed this problem using traditional algorithm selection method~\cite{kashgarani2021algorithm,munoz2013algorithm} or \ac{ML} based ones~\cite{kotthoff2012evaluation}.
Authors in~\cite{hellert2019using}, for example, addressed an adaptive vehicle perception problem in environmental influences.
Performance models are generated based on trained neural networks to predict the suitability of an algorithm to given environmental conditions.  Their goal is to select the best-suited data processing algorithm on-board of an \ac{UAV} under variable circumstances. For a detailed survey on algorithm selection, readers may refer to~\cite{kerschke2019automated}. 
Which algorithms to use is decided on a case-by-case basis and requires domain knowledge. Selecting such a suitable implementation is a challenging task from the developer's perspective. Added to that, complexity increases if we consider several target hardware architectures. Depending on the given situation characterized by available hardware resources and user constraints, the developer has to consider several possible algorithms implementing the same functionality (algorithmic adaptivity) and %
expanding data level parallelism degree across different platforms (parallelism adaptivity).

In the next subsections, we showcase %
this needed of adaptivity through two examples of nowadays applications. We illustrate in an abstract way their complexity to demonstrate the need for parallelism and algorithmic adaptivity and how our approach accounts for this. 
The first application is the \ac{AVP} \footnote{https://www.autoware.org/post/autonomous-valet-parking-2020} from the domain of autonomous driving. This application %
shows how large and complex an application graph may get. The second application is the \ac{ASA}. We motivate the use of mAPN by presenting its dataflow graph in a concise manner. This application is presented in more details, and will be considered in the experimental results in section~\ref{sec:caseStudyASA}.

\subsection{Analysis of AVP Application}
\label{subsec:analysisADAS}

\begin{figure}[h]
\includegraphics[width=\textwidth]{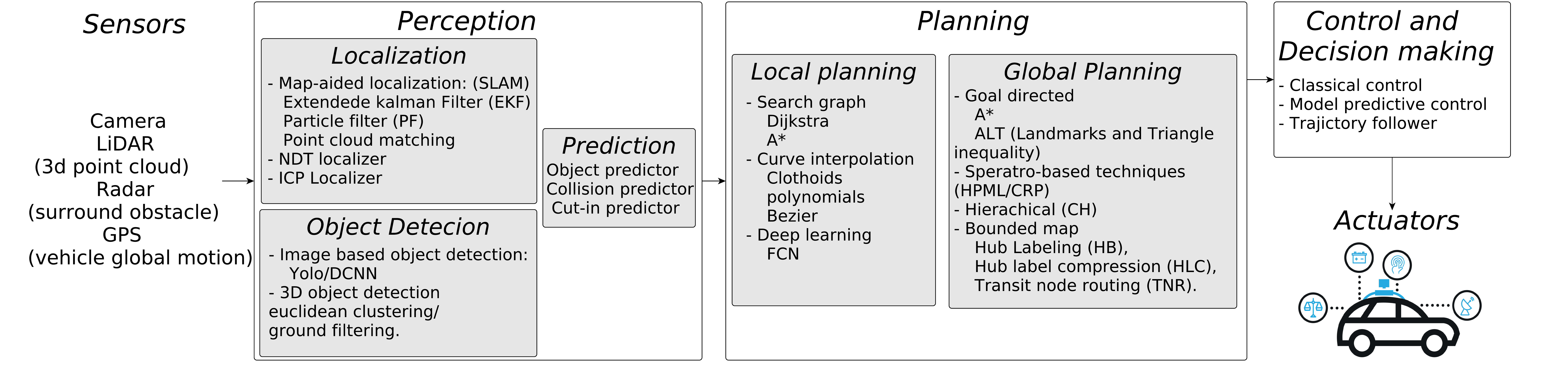}
\caption{Generic autonomous vehicle system}
\label{fig:generic_system}
\end{figure}

A generic autonomous vehicle system is composed of different modules and subsystems implementing several phases that connect the sensors' input used to sense and understand the environment to the actuators executing the driving commands and decision actions (i.e., steering, accelerating, etc.) generated by the system.
Many functionalities are still needed for each module, and various technologies and algorithms are used where no algorithm performs better in all scenarios. The required expertise to develop such complex systems covers several domains such as computer vision, robotics, and vehicle dynamics. A typical \ac{AVP} system is composed of three main phases: the perception phase, the planning phase, and the control or decision making phase.  
For each of these phases more than one possible implementation could be applied and thus be based on different user requirements as well as available hardware. This expands the range of possible implementations for the entire application. Some of these algorithms are stated in~Fig~\ref{fig:generic_system}.

\textbf{Perception:} Data is collected from the sensors to sense and perceive the environment. This data helps the car understand where obstacles are located, determine their positions, and find itself within the whole scene. For this purpose, technologies like cameras, LiDARs, radars, etc., are used. These sensors provide input data to computer vision algorithms to extract useful information. 

\textbf{Planning:}
It involves planning the route to navigate from the current position towards a specific target position, bypassing obstacles. It merges the data from the sensors with the knowledge on the road itself (different obstacles, stop signs, etc.) accumulated during the previous phase to enable the car to plan the path/way to reach the target destination efficiently and safely. 

\textbf{Control/decision making:}
It aims at translating the plan from the previous phase into executable actions (steering, braking, accelerating, etc.) controlling the vehicle. These actions are sent as commands to the hardware/actuators. 

To better understand the complexity of these different phases and the way they are put together to perform the driving task, we will consider the example of the \ac{AVP}. Even though restricted to a specific environment, this driving task still covers many complex scenarios like going to an available parking spot, driving to a pick-up area, stopping for obstacles, etc. Fig.1 in ~\cite{AVP} presents one possible implementation of the AVP. It details how the phases described above apply to the AVP example. For each phase of this application, however, many possible algorithms and implementations might be used depending on the use cases (e.g., parking restricted to AVPs, parking used by both AVPs and normal drivers), the used architecture, and the user constraints. To better grasp the amount of possible implementations, we explore the example of the localization process within the perception phase, which consists of estimating the vehicle's position relative to a reference frame in the environment (global or local map). Mainly, there are three types of localization with different characteristics: absolute localization, relative localization, and odometry~\cite{mohamed2019survey}. Used algorithms depends on the desired scenario. For example odemetry is used under certain conditions such as mapless driving or insufficient map quality for decent relative localization (e.g., tunnel).
Generally, the used inputs and outputs  for these localization types are implementation-specific. Depending on the target use-case, the available sensors, their configuration, accuracy, and limitation, these localization techniques are more or less useful. For example, in a system where GPS/IMU (Inertial Motion Unit) and cameras are available as input sensing devices, combining these input data helps achieving a localization accuracy of 73 cm ~\cite{kuutti2018survey}, which might be enough/acceptable for a specific scenario/application, but not enough for others. Table I in ~\cite{kuutti2018survey} shows 13 possible techniques and implementations for sensor-based localization. Similarly, the localization output serves many functionalities in this application (e.g., behavioral planning, motion control, motion planning, etc.). As such, the need for specific localization algorithms or a combination of algorithms is also dictated by the architecture of the \ac{AVP} system and its intended functionalities. For a more detailed survey on used algorithms in the localization process, the reader may refer to~\cite{kuutti2018survey}.

To summarize, we show that the autonomous driving system is divided into several phases and sub-phases that are meant to interact with each other to realize the driving/parking task. Each of these phases has more than one possible implementation using different techniques. We illustrate the scale at which the number of possible implementations can grow.
The application graph grows even more complex if we consider reflecting parallelism by expanding some nodes to express \ac{DLP}. The choice of the best variant/implementation should depend on the fixed user/application constraints, the application relevant use-cases (dictating the architecture), and the target hardware platform to host the application. Depending on the desired context, the implementation of the application or the expressed parallelism might change. Here comes the need for a compact representation that can express all these variants and reflect the possible algorithmic and parallelism variability. This reduces the effort of building the application to explore all available alternatives to retain those respecting specific constraints and target goals. 
We insist here that the compatibility of used algorithms through the different phases with each other is beyond this paper's scope. Nevertheless, it shall definitely be investigated and considered when building such a multi-alternative representation.

\subsection{Analysis of Automatic Subtitling Application}
\label{subsec:analysisASA}
Another motivational example where more than one implementation serves the same application is the Automatic Subtitling Application (ASA). This example will be 
detailed, compared to the previous one, as it will serve as a case study later in this paper. 
ASA is a complex application that combines three functionalities: Speaker Recognition (SpkR), Speaker Diarization (SD), and Speech Recognition (SpR). These functionalities aim at recognizing who are speaking, when are they speaking, and what are they saying, respectively. 

Automatic subtitling can be found in different scenarios: an offline scenario where a video or audio clip is fed as input and subtitles are generated as an output; or an online scenario (e.g., live TV broadcast) where subtitles are generated in real-time. The second scenario is characterized by harder time constraints, where successfully confirming a person's identity or recognizing the spoken text is as important as getting the answer within a bounded time. If we have to select an adequate implementation, it would not be enough to consider just the accuracy; rather, depending on the user needs, other metrics such as execution time or memory footprint are also important.

Whenever a new ASA has to be implemented for a specific target, there is a general tendency nowadays to compose the solution from existing implementations rather than developing new algorithms. This is related to the \ac{ASP}~\cite{rice1976algorithm} stated in the previous section.  
Depicting the right algorithms to use is decided on a case-by-case basis and depends on variable circumstances (user/hardware constraints, desired scenario, etc.). In the literature, a rich set of algorithms available today to implement the ASA and its different application scenarios~\cite{alvarez2016automating, desplanques2017adaptive, aliprandi2014automatic}.
This subsection reviews the most prominent ones for its different functionalities: SpkR, SD, and SpR.
Afterward, we show in section~\ref{sec:caseStudyASA} how they can be combined to create different ASA variants.

\subsubsection{Speaker Recognition}

\begin{figure}[h]
\centerline{\includegraphics[width=\columnwidth]{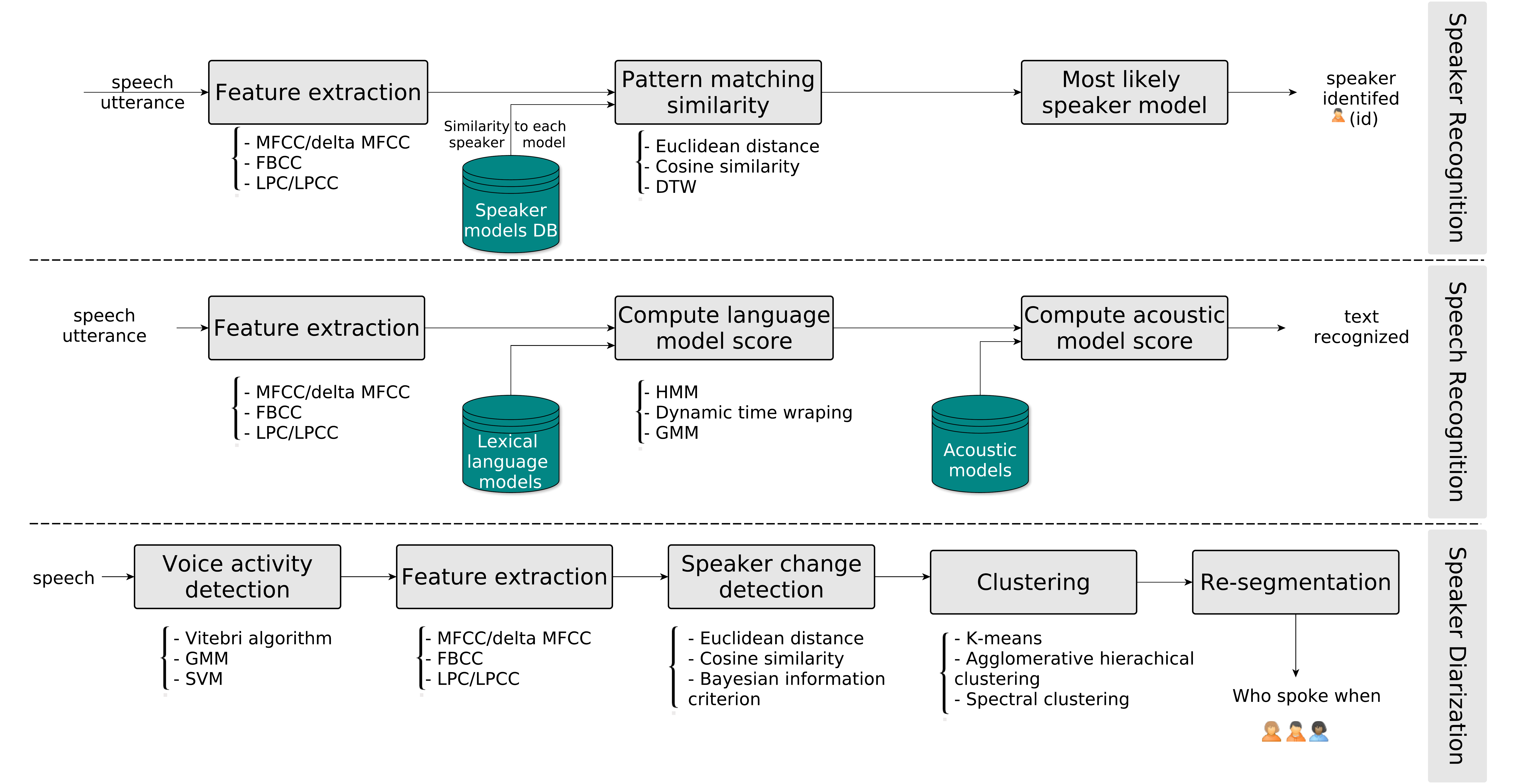}}
\captionsetup{font=footnotesize}
\caption{Speaker Recognition, Speech Recognition and Speaker Diarization functionality graph with possible implementation variants: FE: Mel Frequency Cepstral Coef. (MFCC), Linear Predictive Coef. (LPC), Fourier Bessel Cepstral Ceof. (FBCC). PM: Euclidean Distance (ED), Cosine Similarity (CS), Dynamic Time Wraping(DTW). classification:HMM, DTW, GMM. VAD: Viterbi alg., Gaussian Mixture Models (GMM), Support Vector Machine (SVM). SCD: BIC, ED, CS. Clustering: K-means, AHC, Spectral Clustering. }
\label{all_recognitions}
\end{figure}

SpkR functionality refers to the automated method of identifying or confirming an individual's identity based on their voice. The generic process is described in the upper part of Fig.~\ref{all_recognitions}. At a coarse-grained level, all speaker recognition systems are composed of two main phases: The first phase is Feature Extraction (FE). The features of the speakers are extracted from the voice utterance to create a speaker model. The second phase is pattern matching (PM). It compares the constructed speaker model to all existing models in the database to identify the closest match. Several approaches and algorithms for each phase can be found in the literature. Readers may refer to~\cite{bouraoui2017hardware} for a detailed survey. 
Some of them are presented in Fig.~\ref{all_recognitions}.

\subsubsection{Speaker Diarization}

The goal of SD is to reveal who speaks when. It aims to partition an input audio-stream utterance into segments and annotate each segment with its corresponding label. At a coarse-grained level, most of the SD systems are described as the combination of five phases, as shown in the lower part of Fig.~\ref{all_recognitions}. The ordering of these phases often varies from one system to another \cite{yin2019steps}. The first phase, FE, transforms the audio into acoustic feature vectors (i.e., extracting speaker-specific characteristics). The second phase aims at removing the non-speech regions (e.g., noise or music) by using Voice Activity Detection (VAD) \cite{ghahabi2018robust}. The next phase, Speaker Change Detection (SCD), looks for speaker change points within each speech segment. Then, segments of the same speaker are clustered together \cite{ajmera2003robust}. Finally, speech segments are labeled in the Re-segmentation phase. Its goal is to determine more precisely when each speaker spoke.

\subsubsection{Speech Recognition}

SpR aims at recognizing what is spoken by a speaker. After extracting features from an input utterance, a matching process is applied to identify the corresponding word/sentences being said by comparing the input to a set of words and phonemes models saved in a database (middle part of Fig.~\ref{all_recognitions}). It involves two main steps. The first phase is FE, where the speech signal is transformed into a sequence of pre-phonetic symbols with no linguistic meaning but containing features values. The second step is a classification that includes the acoustic, lexical, and language modeling, which compares the symbols with specific phonetic waveform. To implement these two phases, several well-known algorithms can be found in the literature \cite{yu2016automatic}.

\subsubsection{ASA}

\begin{figure}[t]
\centerline{\includegraphics[width=\columnwidth]{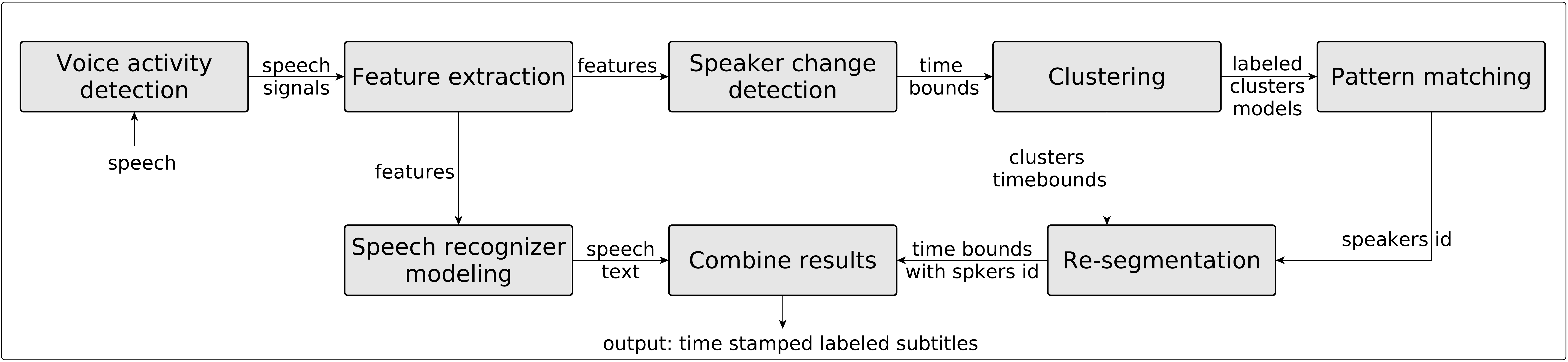}}
\caption{Coarse grained representation of the automatic subtitling application}
\label{fig:coarse_grained}
\end{figure}

From the above description, we can observe that each of the functionalities within ASA (SpkR, SD, SpR) has several implementations using different algorithms. Additionally, they share common phases (e.g., FE) and common algorithms serving different phases (i.e., Classification, PM, and SCD).
 Fig.~\ref{fig:coarse_grained} illustrates a coarse-grained representation of the ASA. Depending on the user/hardware constraints, one may have to specialize the phases for each functionality. Phase reuse and algorithmic choices create a large space of possible variants for ASA as a whole. 

\begin{figure}[t]
\centerline{\includegraphics[width=1\columnwidth]{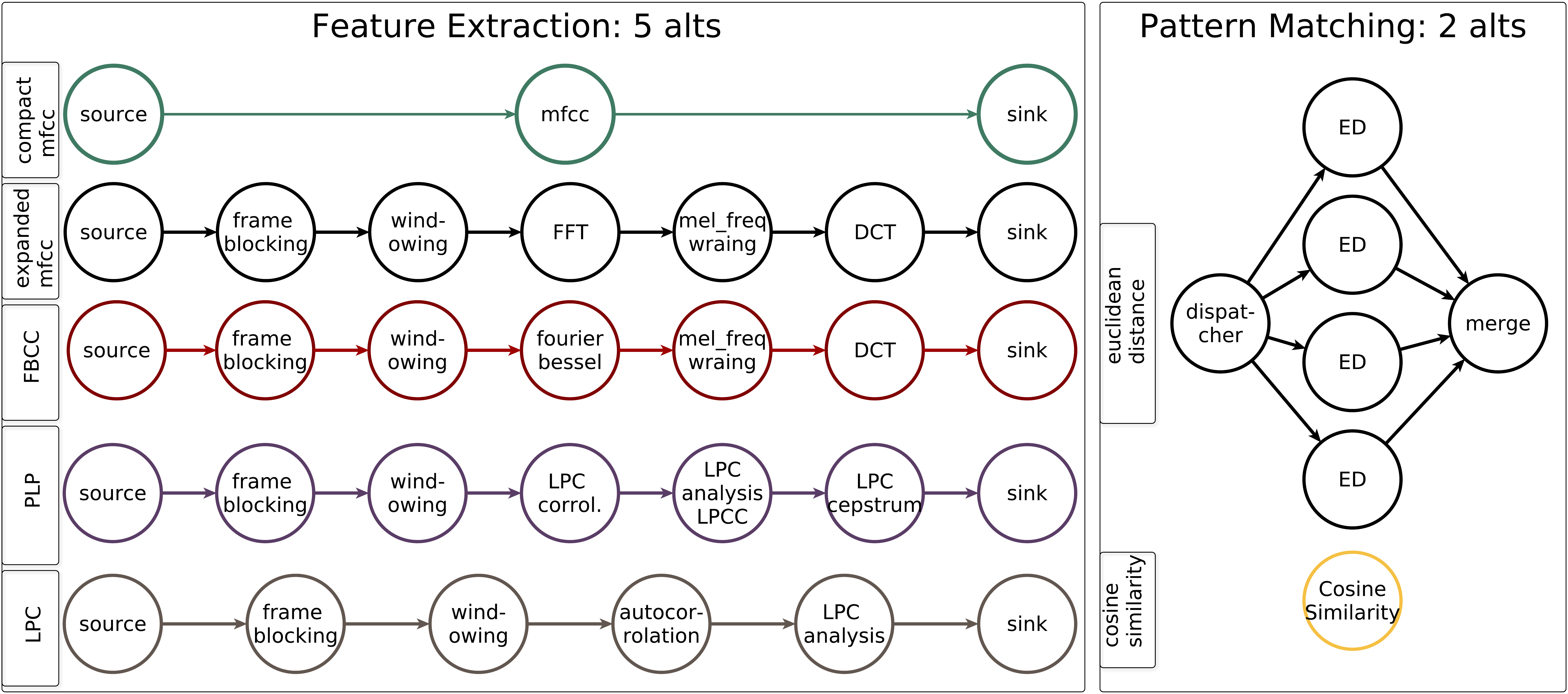}}
\caption{Different feature extraction algorithms }%
\label{fig:FE_alt_SR}
\end{figure}

We detail the FE phase of the SpkR functionality and observe the increasing number of possible variants. 
Speaker characteristics can be categorized based on different features. They are divided into short-term and dynamic features. The first type of features covers those where the length of the frame varies between 20 and 40 ms. The most used algorithms that extract such features are MFCC, FBCC, PLP, or LPC~\cite{tuzun1994comparison, togneri2011overview}. On the other hand, dynamic features describe the time-varying information of utterances (i.e., change of energy). The most used dynamic features are the first derivative and the second derivative of MFCC and LPCC algorithms~\cite{nosratighods2006speaker}. Fig.~\ref{fig:FE_alt_SR} presents five possible implementations of the FE phase (i.e. compact MFCC, expanded MFCC, FBCC, PLP, and LPC). For SpkR, the next phase is the PM, for which we only consider two possible implementations, namely, Euclidean Distance (ED) and Cosine Similarity (CS). 

Using KPNs where computational elements are processes (nodes in the graph) communicating through FIFO channels (edges in the graph), Fig.~\ref{fig:FE_alt_SR} shows a total of 10 different  implementations for SpkR are possible (i.e. MFCC(Compact)+ED/CS, MFCC(Expanded)+ED/CS, PLP+ED/CS, etc.).
For ASA as a whole, many more options are possible. By mixing and matching alternatives from different phases, a big number of variants can be generated. 
In absolute terms, no variant is better than all others. This depends on the fixed user constraints and the available target hardware. 
As mentioned earlier, this application will serve as a case study for our approach. Details of the whole graph and the number of generated variants are presented in section~\ref{sec:caseStudyASA}.

\section{mAPN Formalism}
\label{sec:formalism}
In this section, we introduce a more elaborated version of the mAPN formalism than the one presented in~\cite{bouraoui_et_al:OASIcs.PARMA-DITAM.2021.1}, and give the different definitions of its key components.
We introduce the graph formalism at first, and then discuss how the graph is annotated to allow for fast exploration of application metrics. 
The section closes by comparing our model to the well-known SADF model.

\subsection{Graph Formalism}
Before diving into the graph formalism, we set the terminology to be used and define the mAPN model.
Let $K$ be a KPN model of an application. Considering a subgraph $K'$ of $K$, such that it is itself a KPN with a unique source process and a unique sink process, the designer can \emph{plug} into $K$ an \emph{alternative} KPN, noted $K''$, for $K'$ that starts at the same source process and ends at the same sink process. $K''$ is, therefore, a possible algorithmic substitution of the subgraph $K'$. 
This replacement process for algorithmic adaptivity  can of course be nested.
In this paper, we use colors to graphically distinguish such alternatives.
Additionally, parallel processes within any subgraph are marked in the model to explore parallelism adaptivity. 
During exploration, they are \emph{unfolded} into different alternatives based on the possibilities entered by the designer.
By mixing and matching the different alternatives, one can generate different variants (KPNs of the whole target application).

A KPN is a directed graph composed of concurrent processes (nodes), communicating through unbounded First In First Out (FIFO) channels (edges) having blocking reads and non blocking writes semantics.
Formally, a KPN is a tuple $G=(P, Ch)$, where $P$ is a set of processes, and $Ch \subseteq P \times P$ a set of channels.
An mAPN is a graph that concisely represents many different KPNs. 
We use colors to label and then generate all possible variants. 
Let $\Xi=\{\xi^{1}, \xi^{2},..., \xi^{n}\}$ be the set of colors. 
Similar to KPN, an mAPN is a directed graph composed of processes and channels. 
Unlike KPNs, channels are annotated with colors indicating the alternative that the channel belongs to.
As mentioned before, we graphically distinguish parallel processes with the $\shortparallel$ tag.
Formally,

\begin{definition}%
\label{def:mAPN}
An mAPN graph is a tuple $G=(P, Ch, P^{\shortparallel})$, where:
\begin{itemize}
	\item $P$ is a finite set of processes,
	\item $Ch$ is a finite set of channels, such that $Ch \subseteq P \times P \times \Xi$,
	\item and $P^{\shortparallel} \subset P$, is the subset of parallel processes. 
\end{itemize}  
\end{definition}

\begin{figure}[!t]
\centerline{\includegraphics[width=0.9\columnwidth]{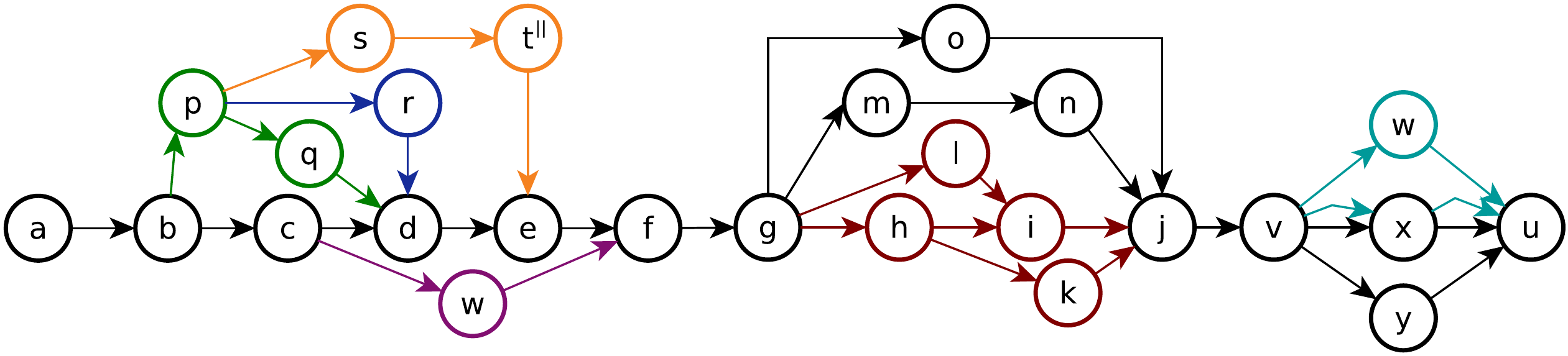}}
\caption{mAPN graph of a synthetic example.}
\label{fig:mAPNExample}
\end{figure}

Let $wr, rd : Ch \rightarrow P$ be two functions that map each channel to the process that writes into it and read from it respectively. 
Let $\widehat{wr}, \widehat{rd} : P \rightarrow \mathcal{P}(Ch)$
be two functions that map each process into a subset of $Ch$ it writes to and reads from respectively (i.e., $\widehat{wr}(p)=\{ch\in Ch, wr(ch)=p\}$). 
We denote by $col$ the function that returns the color of a channel, that is $col: Ch \rightarrow \Xi$. 
The function $\widehat{col}_{rd}$ returns the colors of the channels a process reads from (i.e., $\widehat{col}_{rd}(p)=\cup_{ch \in \widehat{rd}(p)}col(ch)$). 
Analogously, $\widehat{col}_{wr}$ returns the colors of the channels a process writes to. 
In the synthetic mAPN graph example of Fig.~\ref{fig:mAPNExample}, two channels of two different colors ($\turkisline$ and $\blackline$) connect process $v$ to $x$, 
and $\widehat{col}_{wr}(p)=\{\orangeline,\greenline,\blueline\}$.
Process $t$ is parallel, since it is tagged with $\shortparallel$, and thus $t \in P^{\shortparallel}$. 
Unfolding such a process will be detailed in the upcoming section. %
Let $source$ and $sink$ be two functions that return the source and sink processes of a graph. 
Formally, $source(G) = \{p \in P , \widehat{rd}(p) = \emptyset\}$, and $sink(G) = \{p \in P , \widehat{wr}(p)=\emptyset\}$. 
In the mAPN example of Fig.~\ref{fig:mAPNExample}, $source(G)=\{a\}$, 
and  $sink(G)=\{u\}$.

An alternative models how to substitute a piece of the graph with another connected sub-graph implementing the same functionality. 
The channels of the substituting sub-graph have the same color, and the source and sink processes are unique. 
For the substitution to be semantically correct, an alternative cannot fork or join within one subgraph (c.f. formalization as Property~\ref{prop:structeredBlock}).
Formally, 

\begin{definition}%
\label{def:alternative}
A subraph $\alpha(G,\xi)=(P_{\alpha(G,\xi)}, Ch_{\alpha(G,\xi)})$ is an alternative in the mAPN graph $G=(P, Ch, P^{\shortparallel})$, only if:
\begin{itemize}
	\item $\alpha(G,\xi)$ is a connected graph, such that:
		\begin{itemize}
			\item All channels of $Ch$ of color $\xi$ are in $Ch_{\alpha(G,\xi)}$. Formally, $\forall ch \in Ch_{\alpha(G,\xi)}, col(ch)=\xi$ and $\forall ch \in Ch \setminus Ch_{\alpha(G,\xi)}, col(ch) \neq \xi$.
			\item Only processes that are connected to channels of $Ch_{\alpha(G,\xi)}$ are in $P_{\alpha(G,\xi)}$. Formally, $P_{\alpha(G,\xi)} = \bigcup_{ch \in Ch_{\alpha(G,\xi)}} \{wr(ch), rd(ch)\}$.
		\end{itemize}		 
	\item Uniqueness of the source and the sink processes of $\alpha(G,\xi)$. Formally, $|source(\alpha(G,\xi)| = 1$ and $|sink(\alpha(G,\xi)| = 1$. 
\end{itemize}  
\end{definition}

For an alternative $\alpha(G,\xi)$ to be a \emph{potential substitute} of subgraph $\alpha$, both should have the same source and sink processes. 
Consequently, the source process will write channels with at least two different colors, while the sink process will read from channels with at least two different colors.  
For example, the purple (\purpline) subgraph connecting $c$ and $f$ is an alternative to the black (\blackline) subgraph, replacing the functionality of $d$ and $e$ with $w$. The source process $c$ writes to channels with two different colors (\purpline and \blackline), and the sink $f$ also reads from both colors. When considering the black sub-graph that goes through process $c$, the alternatives that starts from node $b$ will not be depicted, and this will avoid any inconsistency with the process $d$ (green and blue alternatives are not considered in this case). 
 
Alternatives can be nested, as is the case of the blue or green (\blueline, \greenline). 
Hence, the flow from $b$ to $d$ can go through $c$ (black (\blackline)), through $p$ and $q$ (green alternative (\greenline)), or through $p$ and $r$ (nested alternative composed of green (\greenline) and blue (\blueline)).  
They are defined as follow:

\begin{definition}%
\label{def:nestedAlternative}
A subraph $\alpha(G)=(P_{\alpha(G)}, Ch_{\alpha(G)})$ is a nested alternative in an mAPN graph $G=(P, Ch, P^{\shortparallel})$, only if:
\begin{itemize}
	\item $\alpha(G)$ is a connected graph.	 
	\item $\alpha(G)$ contains more than one color. Formally, $|\bigcup_{ch \in Ch_{\alpha(G)}} col(ch)| > 1$. 
\end{itemize}
\end{definition}

\emph{End-to-end} alternatives and/or nested alternatives that include $source(G)$ and $sink(G)$ are called \emph{variants}. Formally:

\begin{definition}%
\label{def:variant}
A subgraph $V=(P_{V(G)}, Ch_{V(G)})$ of an mAPN graph $G=(P, Ch, P^{\shortparallel})$ is a variant only if: (i) $V$ is an alternative or a nested alternative, such that $source(V) = source(G)$ and $sink(V) = sink(G)$, and (ii) $V$ is a KPN.   
\end{definition}

We distinguish processes from/at which alternatives fork/join. These processes are important since they identify anchors for mixing and matching alternatives and generating variants. We classify them into the following subsets:
 \begin{itemize}
    \item $\mathcal{F}$: subset of processes of $P$ that writes different channels with different colors. Formally, $\mathcal{F} = \{p \in P, \exists ch_{i}, ch_{j} \in \widehat{wr}(p), i\neq j, col(ch_{i}) \neq col(ch_{j})\}$.
A process in $\mathcal{F}$ is a fork process.
    \item  $\mathcal{J}$ is the subset of processes of $P$ that read different channels with different colors. Formally, $\mathcal{J} = \{p \in P, \exists ch_{i}, ch_{j} \in \widehat{rd}(p), i \neq j, col(ch_{i}) \neq col(ch_{j})\}$.
A process in $\mathcal{J}$ is a join process.
\end{itemize}

In the mAPN of Fig.~\ref{fig:mAPNExample}, $\mathcal{F} = \{b, c, p, v, g\}$ and $\mathcal{J} = \{d, e, f, j, u\}$. 
Based on the collection of alternatives forming the mAPN, one can generate possible variants. 
The generation is based on a set of assumptions that the mAPN is well-formed. 
Formally,  

\begin{definition}[Well-formed mAPN]
\label{wellFormedMAPN}
A well-formed mAPN $G=(P, Ch)$ has these properties:
\begin{enumerate}
 	\item Colors cannot be re-used in disjoint subgraphs. \label{prop:noColorReuse}

   	\item Cycles can only be specified within one same colored subgraph only, that is no alternative can start or end within a cycle. \label{prop:noCycles}

    \item Singularity of the source and sink processes, that is $|source(G)| = |sink(G)| = 1$.\label{prop:singulariy}
       
    \item Preserving KPN semantics: For a process with more than one outgoing color, the number of write channels per color must be the same. Formally, $\forall p \in \mathcal{F}, \forall \xi_{i},\xi_{j} \in \Xi, i \neq j \Rightarrow |\{ch \in \widehat{wr}(p),\xi_{i} \in col(ch)\}| = |\{ch \in \widehat{wr}(p),\xi_{j} \in col(ch)\}|$. 
    Similarly, for a process with more than one incoming color, the number of reading channels per color must be the same. Formally, $\forall p \in \mathcal{J}, \forall \xi_{i},\xi_{j} \in \Xi, i \neq j \Rightarrow |\{ch \in \widehat{rd}(p),\xi_{i} \in col(ch)\}| = |\{ch \in \widehat{rd}(p),\xi_{j} \in col(ch)\}|$. \label{prop:kpnSemantics}

    \item $\forall p \in P^{\shortparallel}, p \notin \mathcal{F} \wedge p \notin \mathcal{J}$.\label{prop:parProcNotInFJ}
    
    \item An alternative and its substituted subgraph form a structured block: a block with a single point of entry and one point of exit.
    Let $\alpha(G)=(P_{\alpha(G)}, Ch_{\alpha(G)})$ be an (nested) alternative of a subraph $\alpha'(G, \xi)=(P_{\alpha'(G,\xi)}, Ch_{\alpha'(G,\xi}))$. The formed block is structured iff none of the processes of $P_{\alpha'(G,\xi)}$ reads or writes channels out of $Ch_{\alpha'(G,\xi)}$. We denote $P_{\alpha'(G,\xi)} \setminus \{souce(P_{\alpha'(G,\xi)}), sink(P_{\alpha'(G,\xi)})\}$ by  $P^{\diamondsuit}$. Formally, \\ 
    $\sum_{p \in P^{\diamondsuit}}|\{ch, ch \in \widehat{wr}(p), col(ch) = \xi\}| 
    = \sum_{p \in P^{\diamondsuit}}|\{ch, ch \in \widehat{rd}(p), col(ch) = \xi\}|$. \label{prop:structeredBlock}
\end{enumerate}
\end{definition}

\begin{figure}[!t]
\centerline{\includegraphics[width=0.5\columnwidth]{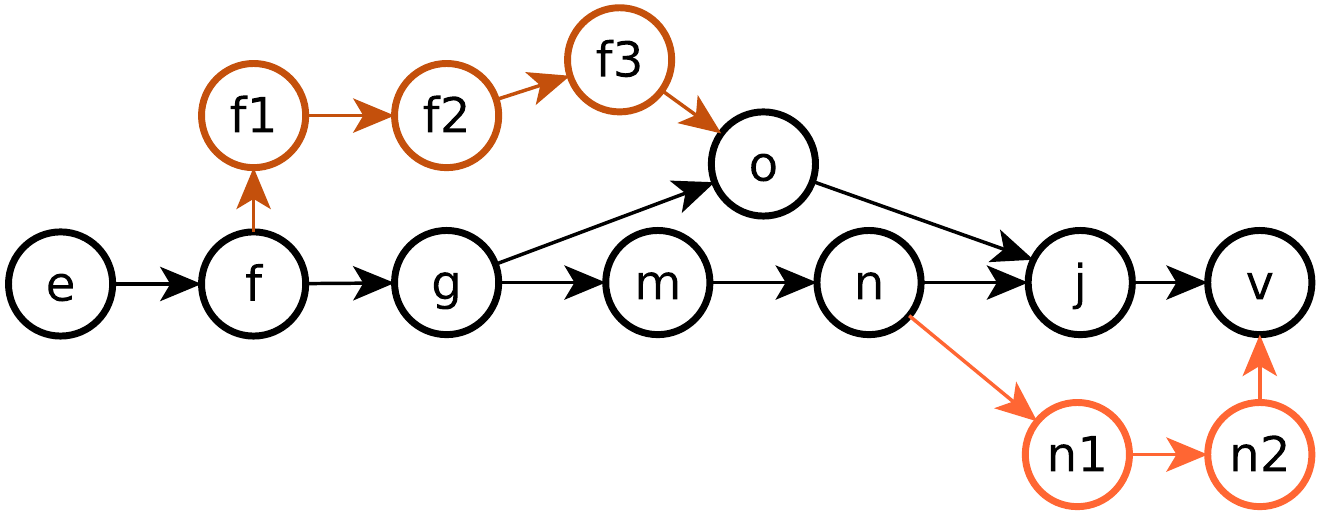}}
\caption{Example of a structured block constraint violation.}
\label{fig:mAPNExample_inconsistency}
\end{figure}

Concretely, properties~(\ref{prop:noColorReuse}) and~(\ref{prop:noCycles}) prevent the designer from adding a subgraph that leads to an inconsistent behavior. 
In addition, property~(\ref{prop:noCycles}) guarantees the semantic independence of the alternatives among each other. Moreover, we simplify the representation of cycles by one process that encompasses the cycle. 
The simplification, however, is an orthogonal problem. 
Property~(\ref{prop:singulariy}) adds a constraint over the uniqueness of the graph source and sink processes. This property is mandatory for the graph exploration (c.f. subsection~\ref{subsec:exploringFeasibleVariants}). In the case of more than one of each, that is $|source(G)| > 1$ (respectively $|sink(G)| > 1$), a unique fictive source (respectively sink) process can be added, making thus the property hold.  
As for preserving KPN semantics (property~(\ref{prop:kpnSemantics})), all fork/join, processes have to write/read, the same number of channels per color.
Parallel processes cannot be a fork or join process (property~(\ref{prop:parProcNotInFJ})). The reason relates to how automatic alternatives unfolding of such processes is performed. 
Finally, property~(\ref{prop:structeredBlock}) prevents from constructing alternatives that forms an inconsistent graph, by observing the structured block constraint.
Fig~\ref{fig:mAPNExample_inconsistency} illustrates such a violation. %
Let's consider the dark orange alternative (formed by the processes $\{f1, f2, f3\}$), 
which is meant to substitute the subgraph formed by $\{f, g, o\}$. 
In this case, the substitution is inconsistent, since channel $(g, m)$ will be removed, 
and therefore process $m$ will fail reading. 
Furthermore, $j$ will wait to read from both channels $(o, j)$ and $(n, j)$, while it is not possible to read from $(n, j)$. 
Similarly, the light orange alternative, formed by the processes $\{n1, n2\}$ is not allowed either. 
The reason is that if the substitution occurs, the channel $(o, j)$ will be removed, and process $o$ will fail writing.
In this illustrated case, we created an inconsistency of the structured black block formed by the processes $\{g, m, o, n, j\}$, either with the dark-orange alternative that substitutes the subgraph formed by the processes $\{f, g, o\}$, or the light orange one that substitutes the subgraph formed by the processes $\{n, j, v\}$ .

\subsection{Annotation of mAPN with Metadata }
\label{subsec:annotations}

An mAPN is enriched with annotations that endow processes with metadata. 
This gives insights that serve to evaluate a variant and check whether it is feasible or not.
Annotations may indicate how a process performs or how robust it is to a distorted input data. 
The developer of the application can define customized metrics to use as well as the way metrics are aggregated during exploration. 
In case of parallel processes, annotations additionally include an enumeration of the parallelism degrees to consider, as well as information on how to derive evaluations for the expanded processes.

\subsubsection{Evaluation in terms of Specific Metric}

Let $M$ be an ordered set of metrics. The values for $m \in M$ are taken from the set of real numbers. 
The reason behind the set $M$ being ordered is to set priorities while choosing among variants (c.f. subsection~\ref{subsec:constraintsChecking}). 
We define a function $\nu(p,m)$ that maps each process $p \in P$ to its evaluation in a given metric $m$. 
If a metric is hardware dependent, $\nu$ returns a vector of values. 
So far, annotations concern only processes, and do not give any information about an alternative. 
For this, we define \emph{operators} for aggregating values, to generalize the evaluation described by annotations to a process network. These operators are metric specific, and are higher order operators. %
We identify two general patterns for aggregating metric values.
The first pattern merges the evaluations coming from input processes. %
Formally, if we consider the set of processes $\{p\} \cup \{p', (p',p, \xi) \in Ch\}$ 
and a metric $m$, then we will apply the merge operator $op_{mg}(m)$ for aggregating the evaluations of processes in $\{p', (p',p, \xi) \in Ch\}$  with regard to $m$. 
The compose pattern, in turn, aggregates the result of $op_{mg}(m)$ with the evaluation of $p$ with regard to $m$. We denote this operator $op_{c}(m)$.
Consequently, $op_{mg}$ is an n-ary higher order operator, while $op_{c}(m)$ is a binary higher order one. 
Arithmetic operators (e.g., $+$, $-$, $*$, $/$, $max$, $min$ and $avr$) are examples of quantitative metrics operators. More elaborated ones can be introduced by the designer and expressed as higher-order functions.

\subsubsection{Annotating Parallel Processes}

As for parallel processes, we define three additional operators: $\rho$, $op^\shortparallel$, and $cost^\shortparallel$.
$\rho$ defines the number of alternatives to drive as well as the \ac{DLP} degree in each one of them. 
$op^\shortparallel$, and $cost^\shortparallel$ operators (c.f. subsection~\ref{subsec:unfolding}), however, define how the metrics evaluations are derived after unfolding parallel processes.

For every element of $P^{\shortparallel}$, $\rho$ returns a set in the power set of $\mathbb{N}$ (formally, $\rho: P^{\shortparallel} \rightarrow \mathcal{P}(\mathbb{N})$). 
$|\rho(p^{\shortparallel})|$ defines the number of alternatives to derive, while each element in $\rho(p^{\shortparallel})$ defines the \ac{DLP} degree. 
Let's consider process $t$ from the synthetic example of Fig.~\ref{fig:mAPNExample} and assume that it has $\rho(t)=\{1,2,4\}$. 
Unfolding such annotation leads to 3 alternatives, as $|\{1,2,4\}|=3$. 
Each alternative has one \ac{DLP} degrees given in the set. 
Process $t$ is therefore duplicated 1 time, 2 times, and 4 times, respectively.

With a few alternatives, it is possible to manually identify and evaluate variants to choose among those that satisfy the constraints.  
For complex applications, however, the number of possible variants can grow high (c.f. ASA case study in section~\ref{sec:caseStudyASA}, where the number of variants reaches 768).
As such, generating variants and evaluating them manually is not practical and prohibitively time consuming for which propose an exploration methodology and tooling (c.f. Section~\ref{sec:methodology}).

\section{mAPN Methodology}
\label{sec:methodology}
As continuity of what established in the previous work~\cite{bouraoui_et_al:OASIcs.PARMA-DITAM.2021.1}, we present in this paper a new methodology in order to explore the graph and extract adequate implementations
Our methodological approach aims to provide designers and developers with a quick way to find an adequate implementation without exhaustively exploring all possible variants of an application. In this section we detail the main components of this methodology, and what they aim for. We present the mAPN tool suite (mAPN$^{TS}$) and the used algorithms for the automatic exploration of the \ac{mAPN} graph, while respecting the constraints introduced by the user. 

\subsection{Overview of mAPN Methodology and the mAPN$^{TS}$}
\begin{figure}[!t]
\centerline{\includegraphics[width=0.99\columnwidth]{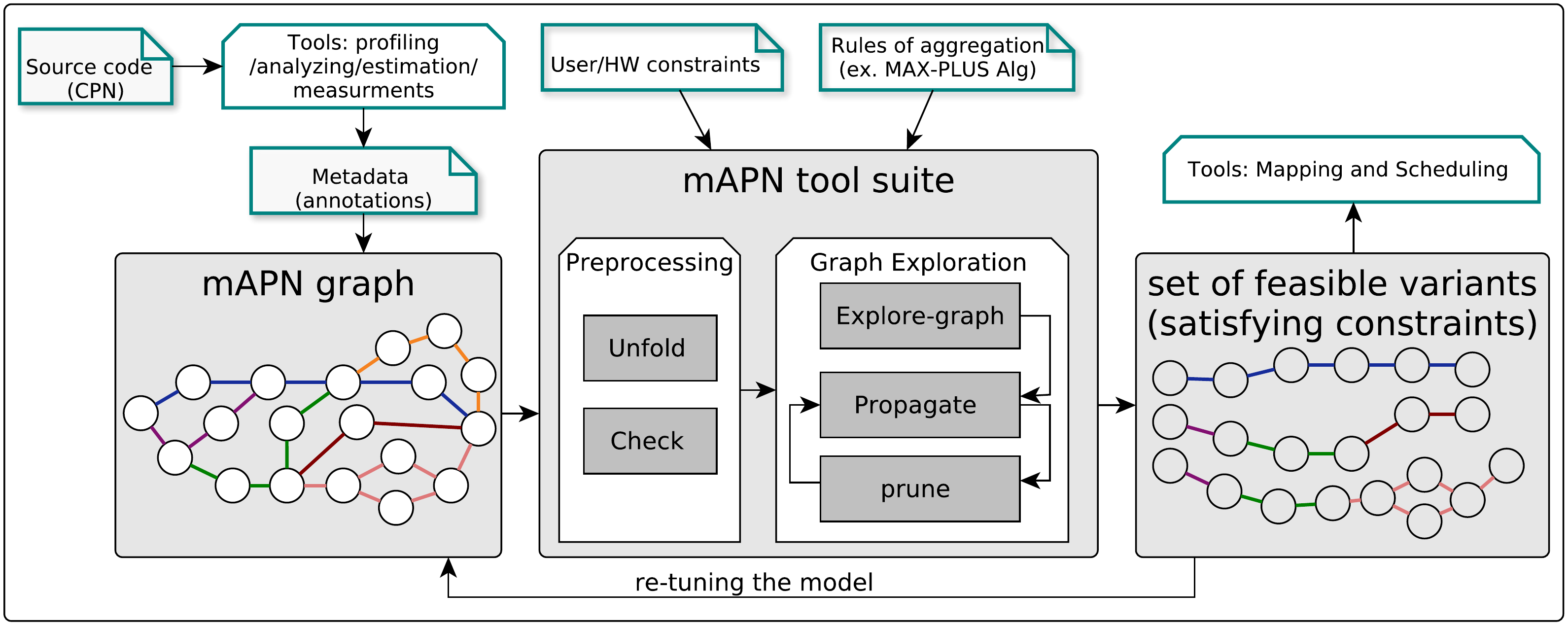}}
\caption{mAPN methodology}
\label{fig:mAPNMethodology}
\end{figure}

The \ac{mAPN} model, its metadata and the design constraints together with the graph exploration are the central components of the methodology (c.f. Fig.~\ref{fig:mAPNMethodology}).

Given an mAPN graph, the individual processes and channels are specified using CPN (C for process network) supported by the SLX tool suite\footnote{https://www.silexica.com/}. 
The metadata (c.f. subsection~\ref{subsec:annotations}) can be collected, for every single process of the mAPN, either manually, based on the designer domain-knowledge; or using available tools. 
For performance estimation, for instance, one can use the estimator of the SLX tool suite, target profiling tools (e.g., the Linux GNU gprof~\footnote{https://sourceware.org/binutils/docs/gprof/}, Score-P\footnote{https://www.vi-hps.org/projects/score-p/}), 
or leverage techniques for Worst-Case Execution Time (WCET) estimation~\cite{lv2009survey}. Moreover, for energy measurements, user can refer to techniques covered by previous research such as PowerPack framework \cite{ge2009powerpack, cameron2005high}. Or by connecting a power measurement devices (i.e., ZES ZIMMER) between the power supply of the inspected system and the
target hardware. For annotations related to memory usage or memory bandwidth Arm MAP\footnote{https://developer.arm.com/documentation/102732/1910/} tool can be used.

In this paper we use actual measurements on the target system for annotating the processes. This way we prevent estimation errors to skew the analysis in Section~\ref{sec:caseStudyASA}.
Constraints set the limiting conditions of an acceptable execution, including resource limitations (e.g., memory size and energy consumption) under which the system should execute to meet its requirements (e.g., deadlines), with the desired level of quality of service (e.g., accuracy).

Since the number of possible variants grows fast, the methodology helps the designer explore the compact mAPN graph and extract feasible variants for the given hardware/user constraints. 
mAPN$^{TS}$ was implemented to this end. 
After unfolding the alternatives of the parallel processes and deriving their individual annotations (c.f. subsection~\ref{subsec:annotations}), 
the tool automatically evaluates variants. This evaluation includes default aggregation rules. The tool suite also allows the user to define customized metrics and corresponding aggregation rules (c.f. subsection~\ref{subsec:rulesAggreg}). 

In the exploration phase (c.f. subsection~\ref{subsec:exploringFeasibleVariants}),  the tool prunes all possible options specified in the compact mAPN graph based on the constraints and outputs a set of possible variants. 
Each of these variants can be further optimized with the rich set of methods for mapping KPN graphs onto multi/many-core platforms~\cite{2013dacSingh, castrillon2011maps}. In~\cite{castrillon2011maps}, the proposed framework provides the possibility to estimate performance modulo the design constraints and offers a set of mapping heuristics. 
In case of an empty set, it is up to the designer to find out how to update the mAPN or engineer solutions to relax the constraints.

\subsection{Graph Unfolding}
\label{subsec:unfolding}

\begin{figure}[!t]
\centerline{\includegraphics[width=0.7\textwidth]{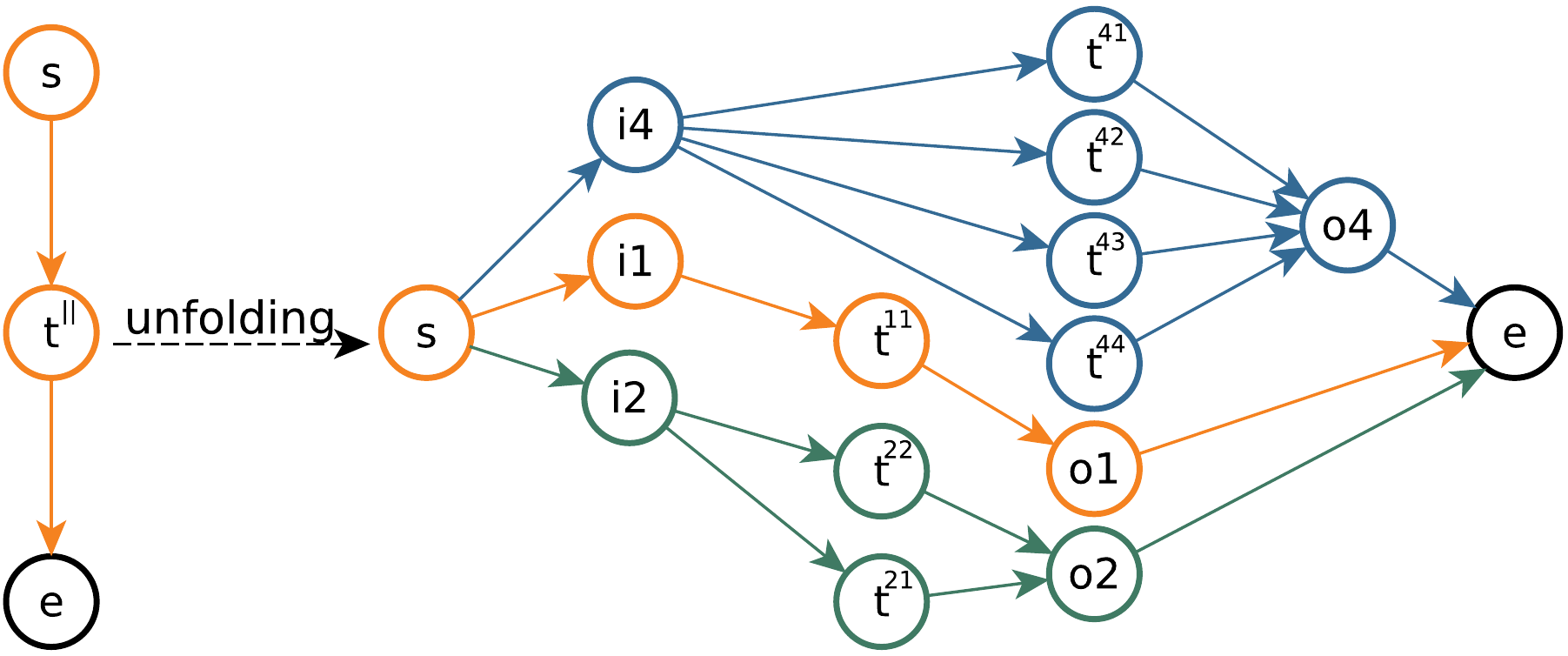}}
\caption{Unfolding parallel process $t$.}
\label{fig:unfoldingDLP}
\end{figure}

As stated before, the compact graph contains some parallel processes that can be duplicated to express \ac{DLP}. The duplication of such nodes leads to additional alternatives expressing parallelism. Fig.~\ref{fig:unfoldingDLP} illustrates this in an example (i.e node $t$). The unfolded node $t$ remains with its same color, while the new unfolded nodes are graphically captured by other colors. 
Unfolding requires adding the intermediary processes that will distribute the workload among the processes and then gather the result. 
In Fig~\ref{fig:unfoldingDLP}, processes $i1$, $i2$, and $i4$ are added as distributing processes, while $o1$, $o2$, and $o4$ are added as gathering processes. 

After unfolding parallel processes, metric evaluations for each of the duplicated processes is needed. To this end, the designer has to define an operator ($op^\shortparallel(m, p, a_i)$) per metric $m$ for computing the evaluations of the duplicated processes ($a_i \in \rho(p)$) from the original process ($p$).
Additionally, the overhead 
due to data distribution and gathering are entered by the designer and specified by a function $overhead^\shortparallel(m,p,a_i)$.

We recall the example of process $t$ in Fig.~\ref{fig:unfoldingDLP} and consider the execution time $e$ as a metric. 
Assuming that there is load balancing in processing the data, $op^\shortparallel(e, t, a_i)=\nu(e,t) / a_i$, that is the execution time is divided among the duplicated processes. 
Consequently, $\nu(e, t^{11}) = \nu(e, t)$, $\nu(e, t^{21}) = \nu(e, t^{22}) = \nu(e, t) / 2$, and $\nu(e, t^{41}) = \nu(e, t^{42}) = \nu(e, t^{43})= \nu(e, t^{44}) = \nu(e, t) / 4$.

\subsection{Rules of Aggregation}
\label{subsec:rulesAggreg}

Using merge and compose operators, a rule aggregates the evaluations from any given metric. 
Formally, the evaluation of a metric $m$ on a variant $V=(P,Ch)$, at a process $p$ (with $p'_{1..n}\in \{p', (p',p) \in Ch\}$) is as follows: 

\[
    \nu(V, m, p)= 
\begin{cases}
    \nu(p,m),& \text{if } p \in source(V)\\
    op_{c}(\nu(p,m), op_{mg}(\nu(V, m, p'_{1}),...,\nu(V, m, p'_{n}))),              & \text{otherwise}
\end{cases}
\]

Recursively, we use the rules above to derive evaluations (with regard to a specific metric) for a larger process network.

We illustrate this calculation through the abstract example of Fig.~\ref{fig:mAPNExample} where we define one metric:  the execution time $e$. 
We focus on the red subgraph, referred as $V_{r}$. For $e$, we assume that $op_{mg}(e) = max$ and $op_{c}(e) = +$ (which boils down to the max-plus algebra traditionally used for timing analysis of dataflow graphs~\cite{baccelli1992synchronization}). %
For clarity reasons, we omit $e$, and we obtain:\\%
$\nu(V_{r},j) =\nu(j)+\nu(g)+ max((\nu(k)+\nu(h)),(\nu(h)+\nu(i)),(\nu(i)+\nu(l)))$
Based on the annotations of processes in terms of specified metric (e.g. execution time), we apply the corresponding aggregation rules (e.gm max-plus algebra) and measure the estimated performance of the whole alternative in terms of this metric. 
Starting from the source node of an mAPN, mAPN$^{TS}$ iteratively aggregates the metrics evaluations at every process of the different variants and for every defined metric.   
The evaluation of a metric at the sink process of a variant is denoted by $\nu(V,m)$.

\subsection{Constraints Checking and Pruning}
\label{subsec:constraintsChecking}

Let $\mathcal{C}$ be a function that maps each metric $m \in M$ to a boolean expression. $\mathcal{C}(m)$ defines the constraint that the variant evaluation should meet with regard to $m$.  
We formally define feasible variants as follows:

\begin{definition}%
\label{def:feasibleVariant}
A variant $V$ is called feasible if and only if all constraints are met, that is: $\forall m \in M, \nu(V, m) \models \mathcal{C}(M)$.
\end{definition}

The number of feasible variants may be large. Our methodology allows the designer to choose the number $b \geq 0$ of the \emph{best} ones to keep. 
Since $M$ is an ordered set, we use prioritization over the metrics to identify the best ones, based on their order.
We call this \emph{pruning}. 
If $b$ is set to $0$, the tool returns the set of all feasible ones. 

\subsection{Exploring Feasible Variants}
\label{subsec:exploringFeasibleVariants}

A previous version of the exploration algorithm was presented in ~\cite{bouraoui_et_al:OASIcs.PARMA-DITAM.2021.1}. That version, besides being recursive, it expands all the alternatives first, and then applies the pruning to extract the feasible alternatives. In this paper, we enhance that exploration method in terms of performance. And in contrast to the exhaustive enumeration of variants from~\cite{bouraoui_et_al:OASIcs.PARMA-DITAM.2021.1} , we propose an incremental algorithm for graph exploration and pruning. 
It explores the different combinations of alternatives in the compact graph itself rather than enumerating all possible variants and then aggregate metric values one by one. 
For the exploration process, all the upcoming algorithms access the following parameters: (i) $G$, an unfolded, well-formed mAPN graph, (ii) $M$, the set of metrics, (iii) $\nu$, the annotations, (iv) $\mathcal{C}$, the constraints, (v) and $b$, the maximum number of (feasible) variants to return.

\texttt{ExploreGraph}, shown in Algorithm~\ref{alg:exploreGraph} is the top level exploration function, which returns a set of feasible variants.%
 The algorithm starts at the source process of the mAPN graph (Line~\ref{alg1line:initSource}) and stops at its sink process, after all processes and channels were explored. 
To each process $pr$ in the graph, a data structure, called $alternatives$, is associated, where we keep track of all discovered alternatives (nested or not) that go from the $source(G)$ towards process $pr$. As such, when the algorithm finishes executing, all discovered (feasible) variants are found in $sink.alternatives$ (Line~\ref{alg1line:returnSink}).

\LinesNumbered

\begin{algorithm}[!t]
	\caption{\textbf{\texttt{ExploreGraph}()}}%
	\label{alg:exploreGraph}
	\SetAlgoLined
	\DontPrintSemicolon	
	\Begin{
		Queue  $Q~~ ; ~~ $int $nextLabel[]$ \;
		$\Xi \leftarrow \emptyset$ \tcc*{Set of colors, initially empty}
		$pr \leftarrow source(G)$  \label{alg1line:initSource} \tcc*{Start at source process}
		\For {each $\xi \in \widehat{col}_{wr}(pr)$}{ \label{alg1line:initQStart}
			$push(Q, \{(pr, \xi)\})$ \tcc*{Initialize Q with pr and its outgoing colors} 
			$\Xi \leftarrow \Xi \cup \{\xi\}$ \tcc*{Add the outgoing colors of pr to the set of colors}
			$push(nextLabel[], \{(\xi, 0)\}$ 
		}\label{alg1line:initQEnd}
		\While { $! Q.empty()$}{ \label{alg1line:outerWhile}
		    $pop(\{(pr, \xi)\}, Q)$ \;
		    Queue $q$ \;
		    \For {each $ch \in \widehat{wr}(pr) \mid col(ch) = \xi$} { \label{alg1line:initqStart}
    			$push(q, ch)$ \tcc*{Push to q all the channels with the current color}
		    } \label{alg1line:initqEnd}
		    \While {$! q.empty()$} {\label{alg1line:innerWhileStart}
			    $pop(ch, q)$\;
			    $pr_{rd} \leftarrow rd(ch)$\;
			    \textbf{\texttt{UpdateExplorationQueue}}$(Q, pr_{rd},\xi,\Xi)$ \label{alg1line:callUpdateExplorationQueue}\tcc*{Update Q while iterating over q}
			    bool $readyToPush \leftarrow \textbf{\texttt{PropagateAlternatives}}(ch, nextLabel[])$ \label{alg1line:callPropagetAlternatives}\;
			    \uIf {$readyToPush = True$}{ 
                    \For {$ch' \in \widehat{wr}(pr) \mid col(ch') = \xi$} {
    			        $push(q, ch')$ \tcc*{continue with upcoming channels of the current color}
                    }
                }
		    }\label{alg1line:innerWhileEnd}
		}\label{alg1line:endOuterWhile}
		\Return $sink(G).alternatives$ \label{alg1line:returnSink} \tcc*{Return all feasible variants}
	}
\end{algorithm}

The exploration queue $Q$ is used to keep track of the processes and colors of their writing channels to explore later. 
The set $\Xi$ and the vector $nextLabel[]$ are used to keep track of the colors to encounter, respectively, the labeling of the number of times a color has been encountered. 
\texttt{ExploreGraph()} starts by initializing the aforementioned variables (Lines~\ref{alg1line:initQStart}-\ref{alg1line:initQEnd}), after which the algorithm  iterates through the graph (outermost \texttt{while} loop, Lines~\ref{alg1line:outerWhile}-\ref{alg1line:endOuterWhile}) until $Q$ becomes empty.
One color $\xi$ is explored at a time. We start by initializing another queue $q$ with the channels of color $\xi$ that $pr$ writes (Lines~\ref{alg1line:initqStart}-\ref{alg1line:initqEnd}). We keep track in $q$ of the channels of the subgraph of the considered color starting from $pr$. 
The inner while loop (Lines~\ref{alg1line:innerWhileStart}-\ref{alg1line:innerWhileEnd}) iterates over the elements in $q$, while updating the exploration queue $Q$ (call to function \texttt{UpdateExplorationQueue()} in line~\ref{alg1line:callUpdateExplorationQueue}).
If we recall the node $j$ in Fig.~\ref{fig:mAPNExample}, and consider the black color, alternatives will not propagate until $j$ receives from all the the previous nodes (i.e. $o$ and $n$). 
Once the process has received from all its predecessors, we propagate the alternative using the function \texttt{PropagateAlternatives} in line~\ref{alg1line:callPropagetAlternatives}. Finally we augment $q$ with the upcoming channels of the considered color (i.e. ($j$, $v$, $black$ in Fig.~\ref{fig:mAPNExample}).

Algorithm~\ref{alg:updateExplorationQueue} (\texttt{UpdateExplorationQueue()}) details when and how the update of the exploration queue $Q$ is performed. 
For this, we associate to each channel the number of times it was visited. 
If at least one of the channels in $\widehat{rd}(pr_{rd})$ have not been visited (lines~\ref{alg2line:forCheckVisitsStart}-\ref{alg2line:forCheckVisitsEnd} derive such information), the queue $Q$ is not updated. Otherwise, we iterate over the colors of the channels in $\widehat{wr}(pr_{rd})$ (for loop between lines~\ref{alg2line:forUpdateStart}-\ref{alg2line:forUpdateEnd}) in order to update $Q$.
The update is performed in two cases. First, when the alternative is newly encountred%
we need to propagate it again through the color $\xi'$  
(Lines~\ref{alg2line:updateCase1Start}-\ref{alg2line:updateCase1End}). 
We recall here Fig~\ref{fig:mAPNExample}. If we explore the red color, assuming that we already finished the black color, and we reach the node $j$, newly learned alternatives from the red path need to be propagated further to the upcoming black. 
Second, whenever we encounter a source process of new alternatives (i.e. $pr_{rd} \in \mathcal{F}$), we need to explore it later (line~\ref{alg2line:updateCase2}). 
Back to Fig~\ref{fig:mAPNExample}, assuming we are exploring the green color and we reach the process $p$ (i.e. fork of new colors), this latter is added to $Q$ for later exploration. 
After the exploration queue $Q$ is updated, the alternatives are propagated from the write process $pr_{wr}$ to the read process $pr_{rd}$ (recall line~\ref{alg1line:callPropagetAlternatives} in Algorithm~\ref{alg:exploreGraph}).

\begin{algorithm}[!t]
\caption{\textbf{\texttt{UpdateExplorationQueue}($Q, pr_{rd},\xi,\Xi$)}}
\label{alg:updateExplorationQueue}

\SetAlgoLined
\DontPrintSemicolon

\Begin
{
    bool $notReceived \leftarrow False$\;
    \For {$ch \in \widehat{rd}(pr_{rd})$} %
    {\label{alg2line:forCheckVisitsStart}
    	\uIf (\tcc*[f]{if $pr_{rd}$ did not receive from $ch$})
    	{$ch.nbVisits = 0$} 
        {
             $notReceived \leftarrow True$\;
        }
    }\label{alg2line:forCheckVisitsEnd}
 
    \uIf{$notReceived = False$}%
    {
        \For 
        {$\xi' \in \widehat{col}_{wr}(pr_{rd})$}
        { \label{alg2line:forUpdateStart}
            \uIf (\tcc*[f]{simple intermediate node for this color})
            {$\xi' =\xi$} 
            {
                $continue$\;
            }
            \uElseIf %
            {$\xi' \in \Xi$} 
            {\label{alg2line:updateCase1Start}
                \If 
                {\textbf{\texttt{HasAlternativeOfColor}}($pr_{rd}$, $\xi'$)}                 				{

                    $push(Q, \{pr_{rd},\xi'\})$ \tcc*[f]{propagate newly learned alts in treated branches}\; 
                }%
            } \label{alg2line:updateCase1End}
            \uElseIf 
            {$\xi' \notin \widehat{col}_{rd}(pr_{rd})$} %
            {\label{alg2line:updateCase2}
                $push(Q, \{pr_{rd},\xi'\})$\;
                $\Xi \leftarrow \Xi \cup \{\xi'\}$\;
            }
            \Else{
                $continue$\;
            }%
    	} \label{alg2line:forUpdateEnd} %
    }%
}   
\end{algorithm}

Algorithm~\ref{alg:propagateAlternatives} details \texttt{PropagateAlternatives()} function. 
It returns a boolean value, $readyToPush$, to indicate whether $pr_{rd}$ is ready to propagate what it has received from its predecessors (after being visited from all of them). 

\begin{algorithm}[!t]
	\caption{\textbf{\texttt{PropagateAlternatives}($ch, nextLabel[]$)}}%
	\label{alg:propagateAlternatives}
	\SetAlgoLined
	\DontPrintSemicolon
\Begin{
	$pr_{rd} \leftarrow rd(ch)~~ ; ~~pr_{wr} \leftarrow wr(ch)~~ ; ~~\xi \leftarrow col(ch) $\; 
	bool $isSink \leftarrow (pr_{rd} = sink(\alpha(G, \xi))) $ \label{alg3line:computeIsSink} \;%
	$ch.nbVisits \leftarrow ch.nbVisits +1$\;
	
    alternative $altsToProp[] \leftarrow \texttt{\textbf{PrepareAltAux}}(pr_{wr}.alternatives, \xi, nextLabel[.])$ \label{alg3line:altsToProp}\; %
    
    \For
    {$alt \in altsToProp[]$ \label{alg3line:forAugmentGraphStart}}
    {
        $alt.AugmentGraph(ch, pr_{wr})$ \tcc*{Add the channel and the read process to the graph}
        $alt.isSink \leftarrow isSink$\;
    } \label{alg3line:forAugmentGraphEnd}%
    $\texttt{\textbf{UpdateAlternatives}}(ch, altsToProp[])$
     \label{alg3line:callUpdateAlternatives} \tcc*{Update: add new alt, merge branches into new alts, update existing alts, compute/update costs} 
    $\texttt{\textbf{Prune}}(altsToProp[], M, \mathcal{C}, b)$ \label{alg3line:callPrune}\;
    bool $clearanceToPropagate \leftarrow True$\;
    \For
    {$ch' \in \widehat{rd}(pr_{rd}) \mid col(ch') = \xi$ \label{alg3line:forClearanceToPropStart}}
    {
        \If{$ch'.nbVisits = 0$}
        {
            $clearanceToPropagate \leftarrow False$\;
            $break$\;
        }%
    }\label{alg3line:forClearanceToPropEnd}%
    
    \Return $clearanceToPropagate$\;
}%
\end{algorithm}
The algorithm starts by checking if $pr_{rd}$ is the sink for an alternative of input color $\xi$ (Line~\ref{alg3line:computeIsSink}). 
From every alternative in $pr_{wr}.alternatives$, we create $altsToProp$, where we put the alternatives of $pr_{wr}$. 
We prepare the alternatives to propagate by assigning corresponding labels and colors, that help later to merge and combine the graphs in process $pr_{rd}$ (line~\ref{alg3line:altsToProp}). 
Each alternative to propagate later via the channel $ch$ is augmented by this new channel and the corresponding process $pr_{rd}$ (loop between Lines~\ref{alg3line:forAugmentGraphStart} and \ref{alg3line:forAugmentGraphEnd}). %
Afterwards, alternatives in $altsToProp$ are propagated and updated (Line~\ref{alg3line:callUpdateAlternatives}). 
This operation starts by creating new alternatives, merging branches into one alternative, or updating some of the existing ones. The $UpdateAlternatives$ function, gets the propagated alternatives (i.e., $altsToProp$) and constructs the new encountered alternatives and updates the existing ones as well as their corresponding costs. Thus it applies the aggregation rules (c.f. subsection~\ref{subsec:rulesAggreg}) to derive the evaluations of the built alternatives. Then comes the pruning function (i.e., $Prune$). Pruning (Line~\ref{alg3line:callPrune}) is performed by applying constraints on the propagated alternative ($altsToProp$) and keeping the $b$ feasible ones (recall that $b$ is the number of alternatives to keep, c.f. Section~\ref{subsec:constraintsChecking}). 

Finally, if $pr_{rd}$ have received from all its predecessors (including $pr_{wr}$ during the present call), it is ready to propagate in its turn (Lines~\ref{alg3line:forClearanceToPropEnd}-\ref{alg3line:forClearanceToPropStart}).

\section{Case Study of Automatic Subtitling Application: mAPN Model}
\label{sec:caseStudyASA}
In this paper, we demonstrate our approach on an \ac{ASA}. Based on the analysis from Section~\ref{subsec:analysisASA} (recalling the graph of Fig.~\ref{fig:coarse_grained}), we conduct a review of existing implementations, commonalities across them and characterize the large design space of algorithmic variants for this particular application. We analyze the fidelity of our model for this use case and evaluate the tooling performance against the state of the art $SDF^3$ tool. 
\subsection{mAPN model for multi-alternative ASA}
\label{subsec:asamAPNModel}
\begin{figure}[!t]
\includegraphics[width=\columnwidth]{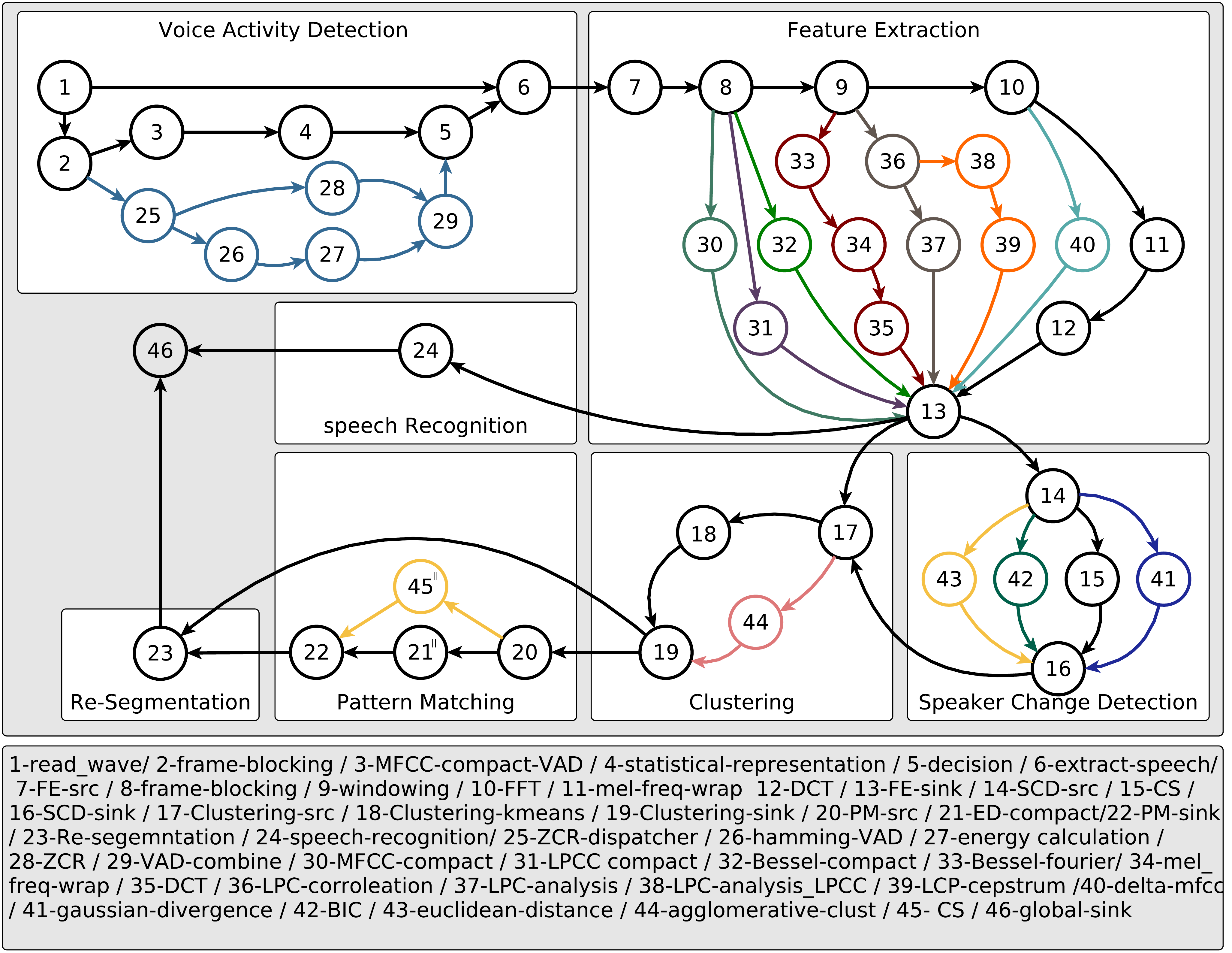}
\setlength\abovecaptionskip{-0.4\baselineskip}
\captionsetup{font=footnotesize}
\caption{Multi-Alternative Process Network for subtitling}
\label{fig:mapn_complex_graph}
\end{figure}

To illustrate the mAPN model, we consider the compact representation for ASA shown in Fig.~\ref{fig:mapn_complex_graph}.
This graph maps to the coarse-grained representation in Fig.~\ref{fig:coarse_grained}. Each phase in the coarse-grained representation is replaced by one or more possible implementations using different colors. For instance, VAD has two different variants, and FE has 8. These two phases alone, give rise to 16 possible variants implementing the part ending at node 13 of the graph. %

So far, we have excluded some of the adaptive parallelism in the variants. To exploit \ac{TLP}, expanded/compacted versions of a PN can be added as an additional algorithmic variant.
The algorithm that implements the MFCC FE can be executed by one PN (i.e., Node 30 ) or expanded over several PNs (4 PNs:8-13). The same is true also for the FE phase's local variants, leading to additional algorithmic variants just for this phase. 
Similarly, \ac{DLP} versions of some phases can be deployed to balance the load across application phases, as seen in~\cite{khasanov2018implicit} or~\cite{schor2014adapnet}.
For example, in the PM phase (node 21), a new alternative can be created where multiple \textit{Euclidean Distance (ED)} nodes run in parallel and then send the results to the \textit{merge} node (recalling the example presented in Fig~\ref{fig:unfoldingDLP}). Every added possibility of implementing a local phase in the compact graph enriches the space of implementation variants. 
After adding some of the discussed \ac{TLP}/\ac{DLP} alternatives,
the number of possible variants grows to 768. 
In fact, the available algorithms in the literature as well as changing circumstances makes this design space ample. Each used algorithm has its own advantages and disadvantages. For few hardware resources available for example, user might extract short-term features to ensure the recognition. However, this might lead to lacking of information. Authors in~\cite{bhatt2021continuous} presented a comparative analysis of some used algorithms for classification and feature extraction. Moreover, pre-processing steps for instance might be used during some circumstances such as noisy environment, where it leads the application to better results such as in ~\cite{istrate2005nist}. In a non-noisy environment, however, users might refer to the alternative that uses a good VAD algorithm to decrease the error rate. If we take the example of the well known problem DIHARD~\cite{ryant2018first} for speaker diarization, authors in ~\cite{wang2021scenario} proposed a scenario-dependent framework for recognizing who spoke when. They presented how the performance of different implementations varies among changing circumstances (i.e. audio-books, meetings, etc). Navigating these rich sets of alternatives requires domain knowledge. 
Based on existing evidence in the literature, the following insights are relevant to help steer the search depending on the desired performance to achieve: 
\begin{itemize}
\item Using too much/too few extracted features leads to overfitting problems/poor accuracy,
\item short-term features are faster to process but might lead to lacking information, 
\item an ASA yields better results when it works on phone conversations rather than broadcasts,
\item a weak VAD output increases the error rate \cite{yin2019steps}, 
\item using a pre-processing step for noise cancellation leads to better results
\item dealing with different audio inputs makes the data different in the recording quality, type of noise, and style of the speech.
\end{itemize}
 
Even with the domain knowledge, finding an adequate implementation while respecting user/HW constraints is nontrivial. Alternatives perform differently depending on the target hardware. Even for specific hardware, it depends on how many other applications are running simultaneously or the mapping of the tasks to the CPU cores (i.e., communication overhead between cores). And for a specific scenario, depending on the changing environment conditions, the achieved performance changes. Thus the need for the mAPN model to ease the process of analyzing different implementations and combine a large number of variants in one compact graph.

\subsection{Model Fidelity Analysis} %
\label{subsec:modelFidelity}
In contrast to the previous work~\cite{bouraoui_et_al:OASIcs.PARMA-DITAM.2021.1}, where only 12 alternatives were evaluated, we enlarge this number in this paper in order to present more concrete results of the fidelity analysis of our approach. To demonstrate the mAPN methodology (Fig.~\ref{fig:mAPNMethodology}), we evaluate the presented alternatives of \ac{ASA} in Fig.~\ref{fig:mapn_complex_graph}. 
All given nodes are processed at least once, while the DLP nodes (21 and 45) are expended to 4 nodes to express parallelism. 

We implemented two alternatives for the VAD phase: the decision based on the zero-crossing rate and energy~\cite{bachu2008separation} (i.e., P 1, 2, 25-29, 5, 6) and the one based on statistical measures~\cite{meduri2012survey} (i.e., PN 1-6). 
In the FE phase, we explore parallelism adaptivity by applying \ac{TLP} to the MFCC and FBCC algorithms (compact versions in nodes 30 and 32, vs. expanded version, 8-13, and 8, 9, 33, 34, 35, 13 respectively), and algorithmic adaptivity by adding other implementations (i.e., LPC, LPCC, FBCC, etc.). 

The SCD phase presents algorithmic adaptivity using four different implementations (i.e., ED, CS, Gaussian divergence, and BIC (Bayesian Information Criteria);. In contrast, the k-means and the agglomerative hierarchical clustering algorithms (18 and 44) are used for the clustering phase. Finally, \ac{DLP} is applied to the PM phase by varying the number of ED-Compact CS-compact (21 and 45) to be expanded to 4. Other degrees of parallelism was unnecessary for this specific application, and we ended up having 224 possible alternatives in total (combinations of 46 different process id).

We explore the space of alternatives to select feasible ones and compare it to a brute force approach, where all implementations are generated and executed. As an evaluation metric, we use the execution time and the aggregation rules described in Section~\ref{subsec:rulesAggreg}. 
For the node-level annotation, we execute all the processes in the target systems and use actual measurements. In a heterogeneous platform, we consider the average of the estimated execution time of processes on different cores. Experiments are performed on a speech of 5 minutes length and considering two platforms: Odroid XU4 (Exynos 5422 big. LITTLE, which has 4 Cortex-A15 and 4 Cortex-A7 cores) and a workstation (GPP) (3.40GHz quad-core Intel(R) Core(TM) i7-6700 CPU). We run the experiments on 4,6, and 8 cores on each platform.

\begin{figure}[!t]
\centerline{\includegraphics[width=1\columnwidth]{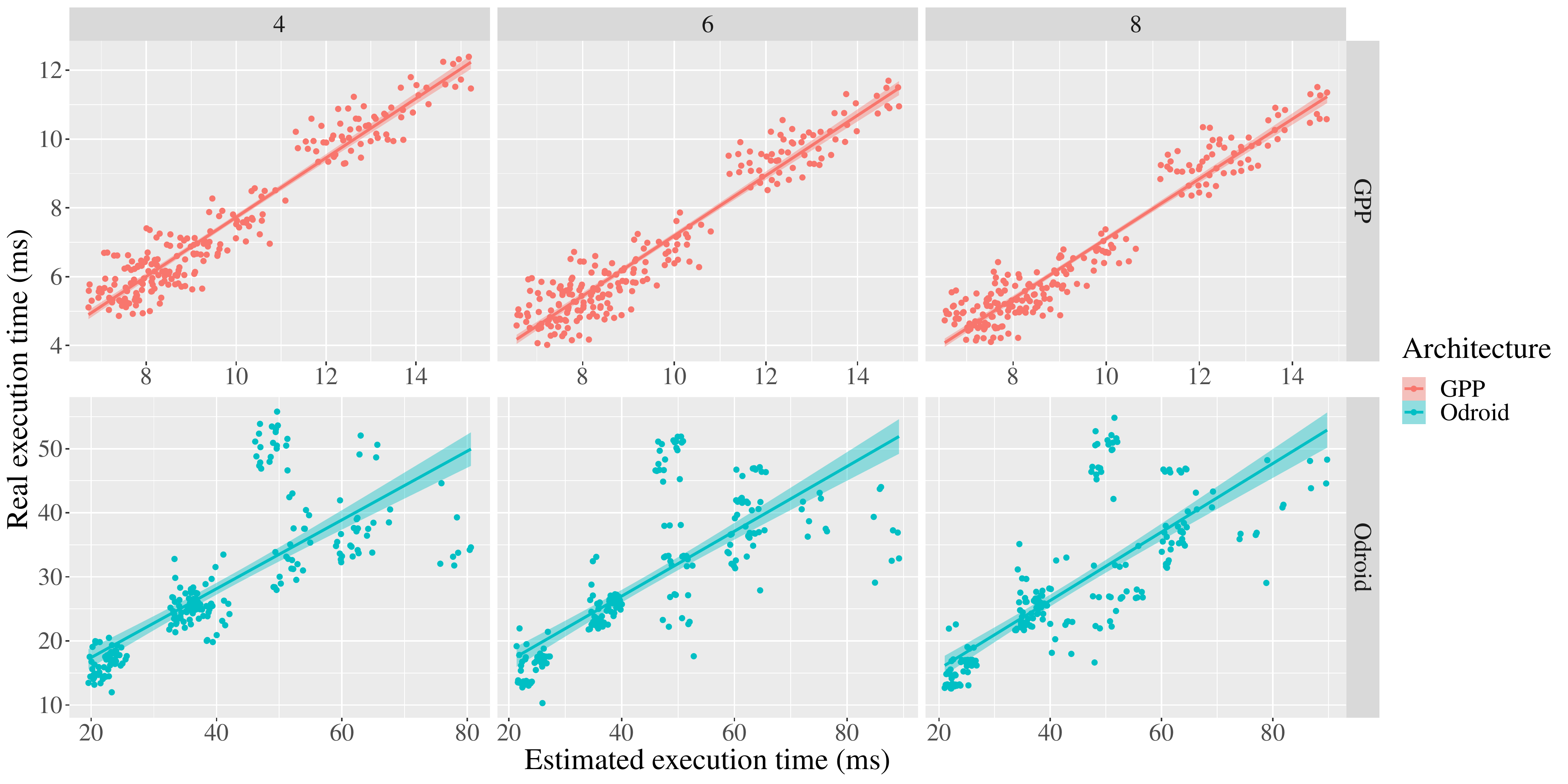}}
\caption{Fidelity analysis of the mAPN methodology for a varying number of cores on GPP/Odroid platforms}
\label{fig:fidelity_analysis}
\end{figure}

The exploration of the space of possibilities, applying the rule of aggregation (c.f. Sec.\ref{subsec:rulesAggreg}) on the evaluation of the node in terms of execution time, gives us an assessment of the 224 considered alternatives. To show the fidelity of our estimation using the mAPN methodology, we plot the correlation between the estimated cost of a variant and the actual execution time of the real implementation.  
We notice that the estimations are generally slower than the real results, which can be explained by the fact that we are not considering the pipelining parallelism hidden in the dataflow graph.
To measure the degree of similarity between the estimation and the real execution time on the platform, we compute the correlation rank. The results are 0.92 and 0.97 using the \textit{pearson's $\rho$} and \textit{Spearman's $\tau$} methods respectively. %
For both methods, very high agreement levels are achieved, proving that our aggregation rules, applied within the context of the mAPN, ensure an acceptable ordering of the alternatives in terms of the considered metric. 
Even with the deviation between the estimations and the measures, we can still say which alternative is better than the other without a time-consuming evaluation of all alternatives on the target hardware. This helps the user decide on the feasible and adequate ones in a large space of variants.

\subsection{Tooling Performance Evaluation}
\label{subsec:toolingPerformance}

The literature counts several MoCs allowing for graph topology changes, as detailed in Section~\ref{sec:relatedWork}. 
SADF is a prominent example of such MoC, where the application topology can be updated by changing the consumption/production rates of tokens. 
In order to put the mAPN methodology into perspective, we evaluate how fast is the decision about the variant to use, compared to SADF. 

To this end, we perform a systematic transformation of mAPN graphs into SADF graphs, where the variants in an mAPN are mimicked with scenarios in an SADF. 
And since mAPN processes do not have consumption and production rates like SADF actors, we solve this mismatch by setting the default value to ``1'' as consumption and production rate in the mapped actors.
An SADF must contain one or more detectors that define the scenarios. An easy way to build an equivalent SADF graph from an mAPN graph is to have one detector that includes all the scenarios (all the alternatives of the mAPN graph). Since it is necessary to have an FSM (or a markov chain), a solution is to iterate between these scenarios concurrently.
In each scenario, the detector will change the consumption/production rates of nodes (0 or 1), and will end up activating one scenario each time.
The generated SADF graph cannot be used to analyze each variant separately since it represents an application that iterates through each alternative. 
We consider several synthetic mAPN graphs while varying the graphs topologies and the number of variants. 
For each of these graphs, we create the equivalent SADF that we feed to $SDF^3$ tool~\cite{stuijk2006sdf} (analysis tool of SADF).
This enables the comparison of the analysis results of each tool and the time needed to provide them.

\begin{figure}[h]
\centerline{\includegraphics[width=\columnwidth]{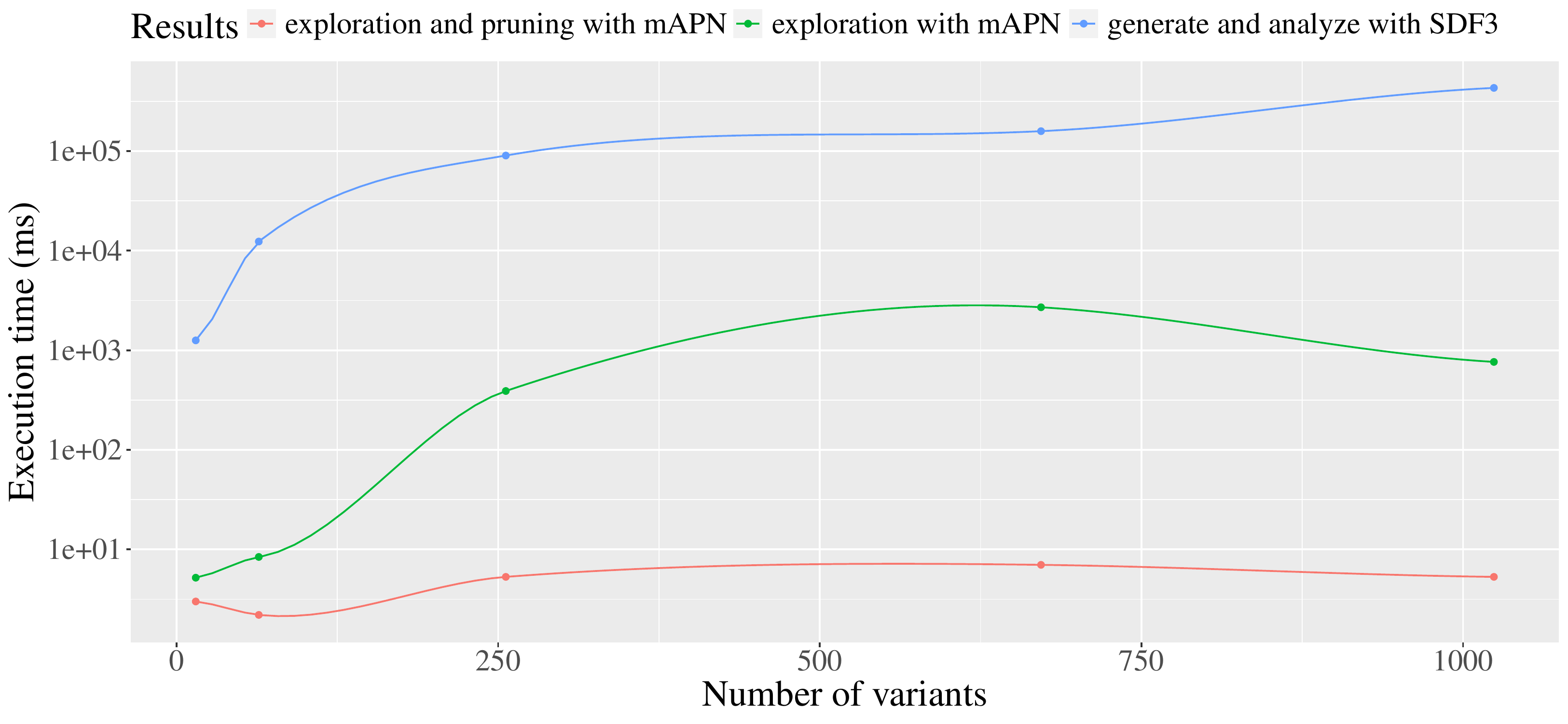}}
\caption{Execution time mAPN tool versus $SDF^3$}
\label{fig:mapn_sdf3}
\end{figure}

Fig.~\ref{fig:mapn_sdf3} compares the execution time for both mAPN$^{TS}$  functionalities and response time analysis using $SDF^3$. 
We compared both tools using random numbers of variants (from 15 to 1024). 
For a large number of variants (1024), it takes more than 7 minutes with $SDF^3$ to analyze all the variants, while mAPN$^{TS}$ takes less than one second. 
And since it is necessary to generate possible variants before analyzing them, the sum of these two steps is represented with the red curve in Fig.~\ref{fig:mapn_sdf3}. 

Besides the fact that the mAPN tool suite performs better in analyzing these variants, it is also possible to prune the graph and get the N best-needed variants in terms of the specified metric (i.e, execution time). Our methodology also allows the exploration of the graph using more than one metric (i.e., execution time, energy). We conducted  experiments on these variants and compared the time needed for the mAPN tool suite to analyze the graph, considering more than one metric. The pruning phase using two metrics takes slightly more time (approximativly 34 ms if we prune and select 5 best variants). 
The exploration and the analysis remain almost the same. Considering the ASA example of 768 variants, analyzing and pruning the graph based on one metric (execution time) takes 2673 ms and 7 ms, respectively, while it takes 2688 ms and 84 ms while considering two metrics (i.e. execution time and energy). For the aggregation rules of these metrics, we used the max-plus algebra for timing analysis and a simple addition for the total energy estimations. 
These additional results proved that the mAPN tool suite is more efficient than the $SDF^3$ in response time.

\section{Related Work: Comparison with Other Models of Computation}
\label{sec:relatedWork}

The literature counts several static and dynamic dataflow models of computation. 

On the one hand, static MoC, such as Synchronous Dataflow~\cite{lee87staticScedDfDsp,1458143} enable static scheduling during compile-time analysis.  
To keep enjoying a number of its formal properties, SDF was extended in different ways. 
Cyclo-Static dataflow (CSDF)~\cite{csdf96}, for example, adds the ability to change the data path in a graph by adding a sequence of possible rates for each port. 
Bui and Lee introduced StreaMorph in~\cite{bui2013streamorph}, an interesting concept of adaptive programs, where the graph structure can morph to adapt to environmental and demand changes. 
The approach was developed with memory usage, energy-saving, and computing resources in mind. 
StreaMorph tackles adaptivity only from the data level parallelism perspective.
PiMM (Parameterized and Interfaced Meta-Model)~\cite{desnos2013pimm} is a meta-model that extends dataflow MoC by adding dynamic reconfigurability and hierarchical capability to the model. 
Applied to SDF, $\pi$SDF~\cite{bhattacharya2001parameterized} is therefore the configurable dynamic MoC. 
It allows parameters to change between iterations, which means that the scenarios are entirely isolated and there is only one active scenario at a time. 
This limits pipeline parallelism capabilities between scenarios. 
Similarly, SADF~\cite{stuijk2011scenario} exhibits the same limitation in its dynamic behavior. 
Being derived from \ac{SDF}, these MoC are bound to rate manipulation and inherit restrictions. 
It is not possible to use SADF model to support analyzing all scenarios at once. However, it is possible with \ac{mAPN} since we distinguish between alternatives at a local node (using colors and labels). For the SADF model we cannot identify different scenarios with local actors since a scenario is defined when the detectors send the control tokens, and each actor has its fixed rates. 

On the other hand, KPNs~\cite{gilles1974semantics} are more expressive. 
They are, however, more challenging to analyze.  
Schor et al. introduced \ac{EPN} in~\cite{schor2014adapnet}, an extension of Process Networks for streaming programming models. 
\ac{EPN} was designed to address the challenge of efficiently exploring task, data, and pipeline parallelism in streaming applications.
For this, they combine several possible granularities in a single specification.
The transformation techniques are replication and unfolding, preventing \ac{EPN}  models from tackling adaptivity from the algorithmic perspective. 
Based on state-of-the-art review, the established investigation, and in accordance 
with~\cite{desnos2013pimm, stuijk2011scenario, rubattu2022pathtracer}, we summarize the MoC comparative 
results in Table~\ref{tab:summaryTable}, in terms of: 
(i) the expressiveness of algorithmic adaptivity, where the value Yes$^{*}$ means that it is possible to express, but in a non-intuitive way, and the value No$^{*}$ means that the formalism does not prevent such a representation, and to the best of our knowledge, no state-of-the art work highlighted the point.  
(ii) the expressiveness of DLP adaptivity, 
(iii) supported annotation metrics (the value - means that it does not apply), 
(iv) variable rate expression, 
(v) the analysis of the variants, 
(vi) and the tool support (the prominent ones).
All of the models listed above allow for the representation of several algorithms for the same application in a single graph at various levels of complexity. 
When it comes to the analyzability of these applications, however, only \ac{mAPN} can support analyzing all the variants simultaneously with the possibility of pruning based on user/hardware constraints. 
In the case of the mentioned models, each variant should be extracted and analyzed aside. In addition, \ac{mAPN} enables the support of several metrics simultaneously. 

\small
\begin{table}[h]
\caption{Characteristics of different MoCs}
\label{tab:summaryTable}
\begin{tabular}{cccccccc}
\hline
  & \begin{tabular}[c]{@{}c@{}}Algorithmic\\Adaptivity\end{tabular} 
  & \begin{tabular}[c]{@{}c@{}}DLP\\Adaptivity\end{tabular}  
  & Annotations
  & \begin{tabular}[c]{@{}c@{}}Variable\\Rates\end{tabular} 
  & \begin{tabular}[c]{@{}c@{}}Variants\\Analysis\end{tabular} 
  & \begin{tabular}[c]{@{}c@{}}Tool\\Support\end{tabular}
  \\ 
\hline
  SDF
  & No
  & No
  & \begin{tabular}[c]{@{}c@{}}Execution\\Time\end{tabular} 
  & No
  &  - 
  & \begin{tabular}[c]{@{}c@{}} $SDF^3$~\cite{stuijk2006sdf}/\\ Ptolemy II~\cite{lee1998framework} \end{tabular}                                                                  
  \\ 
\hline
  CSDF 
  & No$^{*}$
  & No$^{*}$
  & \begin{tabular}[c]{@{}c@{}}Execution\\Time\end{tabular} 
  & Yes/No
  & \begin{tabular}[c]{@{}c@{}}Each one\\  apart\end{tabular}   
  & $SDF^3$~\cite{stuijk2006sdf}    
  \\
\hline
  \begin{tabular}[c]{@{}c@{}}Strea- \\ Morph \end{tabular} 
  & No
  & Yes
  & - 
  & Yes
  & \begin{tabular}[c]{@{}c@{}}Each one\\apart\end{tabular}    
  & -%
  \\
\hline
  PiMM    
  & Yes$^{*}$
  & No$^{*}$
  & \begin{tabular}[c]{@{}c@{}}Execution \\ Time\end{tabular} 
  & Yes                                                                                                                    
  & \begin{tabular}[c]{@{}c@{}}Each one \\ apart\end{tabular}   
  & \begin{tabular}[c]{@{}c@{}}Preesm/\\ Spider~\cite{Preesm}\end{tabular}
  \\
\hline
  SADF 
  & Yes$^{*}$
  & No$^{*}$
  & \begin{tabular}[c]{@{}c@{}}Execution \\ Time\end{tabular} 
  & Yes
  & \begin{tabular}[c]{@{}c@{}}Each one\\ apart\end{tabular}    
  & $SDF^3$~\cite{stuijk2006sdf}
  \\
\hline
  KPN  
  & No
  & No     
  & -
  & -
  & \begin{tabular}[c]{@{}c@{}}Each one \\ apart\end{tabular}   
  & \begin{tabular}[c]{@{}c@{}} Ptolemy II\\ \cite{lee1998framework}\end{tabular}   
  \\
\hline
  EPN  
  & No
  & Yes   
  & -
  & -
  &    
  & AdaPNet~\cite{Schor2014Oct12} 
  \\
\hline
  mAPN  
  & Yes
  & Yes                                            
  & Multiple                                                  
  & - 
  & \begin{tabular}[c]{@{}c@{}}Compact\\  graph\end{tabular}   
  &  mAPN$^{TS}$
  \\
\hline
\end{tabular}
\end{table}                   
         
\normalsize

\section{Conclusion and Future Work}
\label{sec:conclusion}
In this paper, we presented \ac{mAPN}, a new high-level model, where multiple variants exists in the same graph. In particular, we showed how the number of variants grows fast and how complex the dataflow graph could be. The presented methodology allowed us to concisely express these several algorithmic variants in a single compact graph while supporting algorithmic and parallelism adaptivity. 
We presented how \ac{mAPN} is enriched with additional meta-data (i.e., rules of aggregations, annotations in terms of specific metric), which enlarges the variant space and eases the process of extracting feasible variants while meeting user and hardware constraints. Our model also allows for customized aggregation rules of user-defined metrics.
We motivated the \ac{mAPN} approach with 224 variants of the \ac{ASA} example. We used aggregation methods to analyze the end-to-end paths of the co-existing algorithmic variants in this compact graph. 
We analyzed these evaluations in terms of the defined execution time metric. We also showed that we performs better than the well-known tool $SDF^3$ in terms of analysis assessment. 
The \ac{mAPN} tool is implemented using Python3 as a proof of concept. An implementation in C++ (like $SDF^3$) would perform surely even faster.
The presented \ac{mAPN} model allowed us to reason about these metrics and algorithmic adaptivity concisely while achieving the most suitable implementations in terms of the addressed user and hardware constraints.

In future work, we will address questions of supporting additional analysis capabilities of an \ac{mAPN} graph while considering the methods for mapping graphs onto heterogeneous multicores. 
We are interested in adding algorithmic variants to the design space exploration for application adaptivity in hybrid mapping methodologies as proposed in~\cite{goens2017tetris,khasanov_cases21}.
We will also investigate other aggregation rules over more abstract domain-specific metrics such as safety. And finally, additional future work would be carried out to let mAPN methodology support a suitable switching mechanism and adaptability at run-time.

\bibliographystyle{ACM-Reference-Format}
\bibliography{references}


\begin{thebibliography}{60}


\ifx \showCODEN    \undefined \def \showCODEN     #1{\unskip}     \fi
\ifx \showDOI      \undefined \def \showDOI       #1{#1}\fi
\ifx \showISBNx    \undefined \def \showISBNx     #1{\unskip}     \fi
\ifx \showISBNxiii \undefined \def \showISBNxiii  #1{\unskip}     \fi
\ifx \showISSN     \undefined \def \showISSN      #1{\unskip}     \fi
\ifx \showLCCN     \undefined \def \showLCCN      #1{\unskip}     \fi
\ifx \shownote     \undefined \def \shownote      #1{#1}          \fi
\ifx \showarticletitle \undefined \def \showarticletitle #1{#1}   \fi
\ifx \showURL      \undefined \def \showURL       {\relax}        \fi
\providecommand\bibfield[2]{#2}
\providecommand\bibinfo[2]{#2}
\providecommand\natexlab[1]{#1}
\providecommand\showeprint[2][]{arXiv:#2}

\bibitem[\protect\citeauthoryear{??}{AVP}{[n.d.]}]%
        {AVP}
 \bibinfo{year}{[n.d.]}\natexlab{}.
\newblock \bibinfo{title}{Autonomous Valet Parking 2020}.
\newblock
  \bibinfo{howpublished}{\url{https://www.autoware.org/post/autonomous-valet-parking-2020}}.
\newblock
\newblock
\shownote{Accessed: 2022-02-21.}


\bibitem[\protect\citeauthoryear{Ajmera and Wooters}{Ajmera and
  Wooters}{2003}]%
        {ajmera2003robust}
\bibfield{author}{\bibinfo{person}{Jitendra Ajmera} {and}
  \bibinfo{person}{Chuck Wooters}.} \bibinfo{year}{2003}\natexlab{}.
\newblock \showarticletitle{A robust speaker clustering algorithm}. In
  \bibinfo{booktitle}{\emph{2003 IEEE Workshop on Automatic Speech Recognition
  and Understanding (IEEE Cat. No. 03EX721)}}. IEEE, \bibinfo{pages}{411--416}.
\newblock


\bibitem[\protect\citeauthoryear{Aliprandi, Scudellari, Gallucci, Piccinini,
  Raffaelli, del Pozo, Alvarez, Arzelus, Cassaca, Luis,
  et~al\mbox{.}}{Aliprandi et~al\mbox{.}}{2014}]%
        {aliprandi2014automatic}
\bibfield{author}{\bibinfo{person}{C. Aliprandi}, \bibinfo{person}{C.
  Scudellari}, \bibinfo{person}{I. Gallucci}, \bibinfo{person}{N. Piccinini},
  \bibinfo{person}{M. Raffaelli}, \bibinfo{person}{A. del Pozo},
  \bibinfo{person}{A. Alvarez}, \bibinfo{person}{Ha. Arzelus},
  \bibinfo{person}{R. Cassaca}, \bibinfo{person}{T. Luis}, {et~al\mbox{.}}}
  \bibinfo{year}{2014}\natexlab{}.
\newblock \showarticletitle{Automatic Live Subtitling: state of the art,
  expectations and current trends}. In \bibinfo{booktitle}{\emph{Proceedings of
  NAB Broadcast Engineering Conference: Papers on Advanced Media Technologies,
  Las Vegas}}. \bibinfo{pages}{23}.
\newblock


\bibitem[\protect\citeauthoryear{{\'A}lvarez, Mendes, Raffaelli, Lu{\'\i}s,
  Paulo, Piccinini, Arzelus, Neto, Aliprandi, and Del~Pozo}{{\'A}lvarez
  et~al\mbox{.}}{2016}]%
        {alvarez2016automating}
\bibfield{author}{\bibinfo{person}{Aitor {\'A}lvarez}, \bibinfo{person}{Carlos
  Mendes}, \bibinfo{person}{Matteo Raffaelli}, \bibinfo{person}{Tiago
  Lu{\'\i}s}, \bibinfo{person}{S{\'e}rgio Paulo}, \bibinfo{person}{Nicola
  Piccinini}, \bibinfo{person}{Haritz Arzelus}, \bibinfo{person}{Jo{\~a}o
  Neto}, \bibinfo{person}{Carlo Aliprandi}, {and} \bibinfo{person}{Arantza
  Del~Pozo}.} \bibinfo{year}{2016}\natexlab{}.
\newblock \showarticletitle{Automating live and batch subtitling of multimedia
  contents for several European languages}.
\newblock \bibinfo{journal}{\emph{Multimedia Tools and Applications}}
  \bibinfo{volume}{75}, \bibinfo{number}{18} (\bibinfo{year}{2016}),
  \bibinfo{pages}{10823--10853}.
\newblock


\bibitem[\protect\citeauthoryear{Ansel, Chan, Wong, Olszewski, Zhao, Edelman,
  and Amarasinghe}{Ansel et~al\mbox{.}}{2009}]%
        {ansel2009petabricks}
\bibfield{author}{\bibinfo{person}{Jason Ansel}, \bibinfo{person}{Cy Chan},
  \bibinfo{person}{Yee~Lok Wong}, \bibinfo{person}{Marek Olszewski},
  \bibinfo{person}{Qin Zhao}, \bibinfo{person}{Alan Edelman}, {and}
  \bibinfo{person}{Saman Amarasinghe}.} \bibinfo{year}{2009}\natexlab{}.
\newblock \showarticletitle{PetaBricks: a language and compiler for algorithmic
  choice}.
\newblock \bibinfo{journal}{\emph{ACM Sigplan Notices}} \bibinfo{volume}{44},
  \bibinfo{number}{6} (\bibinfo{year}{2009}), \bibinfo{pages}{38--49}.
\newblock


\bibitem[\protect\citeauthoryear{Baccelli, Cohen, Olsder, and Quadrat}{Baccelli
  et~al\mbox{.}}{1992}]%
        {baccelli1992synchronization}
\bibfield{author}{\bibinfo{person}{Fran{\c{c}}ois Baccelli},
  \bibinfo{person}{Guy Cohen}, \bibinfo{person}{Geert~Jan Olsder}, {and}
  \bibinfo{person}{Jean-Pierre Quadrat}.} \bibinfo{year}{1992}\natexlab{}.
\newblock \showarticletitle{Synchronization and linearity: an algebra for
  discrete event systems}.
\newblock  (\bibinfo{year}{1992}).
\newblock


\bibitem[\protect\citeauthoryear{Bachu, Kopparthi, Adapa, and Barkana}{Bachu
  et~al\mbox{.}}{2008}]%
        {bachu2008separation}
\bibfield{author}{\bibinfo{person}{RG Bachu}, \bibinfo{person}{S Kopparthi},
  \bibinfo{person}{B Adapa}, {and} \bibinfo{person}{BD Barkana}.}
  \bibinfo{year}{2008}\natexlab{}.
\newblock \showarticletitle{Separation of voiced and unvoiced using zero
  crossing rate and energy of the speech signal}. In
  \bibinfo{booktitle}{\emph{American Society for Engineering Education (ASEE)
  zone conference proceedings}}. American Society for Engineering Education,
  \bibinfo{pages}{1--7}.
\newblock


\bibitem[\protect\citeauthoryear{Bergenhem, Shladover, Coelingh, Englund, and
  Tsugawa}{Bergenhem et~al\mbox{.}}{2012}]%
        {bergenhem2012overview}
\bibfield{author}{\bibinfo{person}{Carl Bergenhem}, \bibinfo{person}{Steven
  Shladover}, \bibinfo{person}{Erik Coelingh}, \bibinfo{person}{Christoffer
  Englund}, {and} \bibinfo{person}{Sadayuki Tsugawa}.}
  \bibinfo{year}{2012}\natexlab{}.
\newblock \showarticletitle{Overview of platooning systems}. In
  \bibinfo{booktitle}{\emph{Proceedings of the 19th ITS World Congress, Oct
  22-26, Vienna, Austria (2012)}}.
\newblock


\bibitem[\protect\citeauthoryear{Bhatt, Jain, and Dev}{Bhatt
  et~al\mbox{.}}{2021}]%
        {bhatt2021continuous}
\bibfield{author}{\bibinfo{person}{Shobha Bhatt}, \bibinfo{person}{Anurag
  Jain}, {and} \bibinfo{person}{Amita Dev}.} \bibinfo{year}{2021}\natexlab{}.
\newblock \showarticletitle{Continuous speech recognition technologies—A
  review}.
\newblock \bibinfo{journal}{\emph{Recent developments in acoustics}}
  (\bibinfo{year}{2021}), \bibinfo{pages}{85--94}.
\newblock


\bibitem[\protect\citeauthoryear{Bhattacharya and Bhattacharyya}{Bhattacharya
  and Bhattacharyya}{2001}]%
        {bhattacharya2001parameterized}
\bibfield{author}{\bibinfo{person}{Bishnupriya Bhattacharya} {and}
  \bibinfo{person}{Shuvra~S Bhattacharyya}.} \bibinfo{year}{2001}\natexlab{}.
\newblock \showarticletitle{Parameterized dataflow modeling for DSP systems}.
\newblock \bibinfo{journal}{\emph{IEEE Transactions on Signal Processing}}
  \bibinfo{volume}{49}, \bibinfo{number}{10} (\bibinfo{year}{2001}),
  \bibinfo{pages}{2408--2421}.
\newblock


\bibitem[\protect\citeauthoryear{{Bilsen}, {Engels}, {Lauwereins}, and
  {Peperstraete}}{{Bilsen} et~al\mbox{.}}{1996}]%
        {csdf96}
\bibfield{author}{\bibinfo{person}{G. {Bilsen}}, \bibinfo{person}{M. {Engels}},
  \bibinfo{person}{R. {Lauwereins}}, {and} \bibinfo{person}{J.
  {Peperstraete}}.} \bibinfo{year}{1996}\natexlab{}.
\newblock \showarticletitle{Cycle-static dataflow}.
\newblock \bibinfo{journal}{\emph{IEEE Transactions on Signal Processing}}
  \bibinfo{volume}{44}, \bibinfo{number}{2} (\bibinfo{year}{1996}),
  \bibinfo{pages}{397--408}.
\newblock


\bibitem[\protect\citeauthoryear{Bouraoui, Jerad, and Castrillon}{Bouraoui
  et~al\mbox{.}}{2021}]%
        {bouraoui_et_al:OASIcs.PARMA-DITAM.2021.1}
\bibfield{author}{\bibinfo{person}{Hasna Bouraoui}, \bibinfo{person}{Chadlia
  Jerad}, {and} \bibinfo{person}{Jeronimo Castrillon}.}
  \bibinfo{year}{2021}\natexlab{}.
\newblock \showarticletitle{{Towards Adaptive Multi-Alternative Process
  Network}}. In \bibinfo{booktitle}{\emph{12th Workshop on Parallel Programming
  and Run-Time Management Techniques for Many-core Architectures and 10th
  Workshop on Design Tools and Architectures for Multicore Embedded Computing
  Platforms (PARMA-DITAM 2021)}} \emph{(\bibinfo{series}{Open Access Series in
  Informatics (OASIcs)})}, \bibfield{editor}{\bibinfo{person}{Jo\~{a}o Bispo},
  \bibinfo{person}{Stefano Cherubin}, {and} \bibinfo{person}{Jos\'{e} Flich}}
  (Eds.), Vol.~\bibinfo{volume}{88}. \bibinfo{publisher}{Schloss Dagstuhl --
  Leibniz-Zentrum f{\"u}r Informatik}, \bibinfo{address}{Dagstuhl, Germany},
  \bibinfo{pages}{1:1--1:11}.
\newblock
\showISBNx{978-3-95977-181-8}
\showISSN{2190-6807}
\urldef\tempurl%
\url{https://doi.org/10.4230/OASIcs.PARMA-DITAM.2021.1}
\showDOI{\tempurl}


\bibitem[\protect\citeauthoryear{Bouraoui, Jerad, Chattopadhyay, and
  Hadj-Alouane}{Bouraoui et~al\mbox{.}}{2017}]%
        {bouraoui2017hardware}
\bibfield{author}{\bibinfo{person}{Hasna Bouraoui}, \bibinfo{person}{Chadlia
  Jerad}, \bibinfo{person}{Anupam Chattopadhyay}, {and}
  \bibinfo{person}{Nejib~Ben Hadj-Alouane}.} \bibinfo{year}{2017}\natexlab{}.
\newblock \showarticletitle{Hardware architectures for embedded speaker
  recognition applications: a survey}.
\newblock \bibinfo{journal}{\emph{ACM Transactions on Embedded Computing
  Systems (TECS)}} \bibinfo{volume}{16}, \bibinfo{number}{3}
  (\bibinfo{year}{2017}), \bibinfo{pages}{78}.
\newblock


\bibitem[\protect\citeauthoryear{Bui and Lee}{Bui and Lee}{2013}]%
        {bui2013streamorph}
\bibfield{author}{\bibinfo{person}{Dai Bui} {and} \bibinfo{person}{Edward~A
  Lee}.} \bibinfo{year}{2013}\natexlab{}.
\newblock \showarticletitle{StreaMorph: a case for synthesizing
  energy-efficient adaptive programs using high-level abstractions}. In
  \bibinfo{booktitle}{\emph{Proceedings of EMSOFT}}. IEEE Press,
  \bibinfo{pages}{20}.
\newblock


\bibitem[\protect\citeauthoryear{Cameron, Ge, and Feng}{Cameron
  et~al\mbox{.}}{2005}]%
        {cameron2005high}
\bibfield{author}{\bibinfo{person}{Kirk~W Cameron}, \bibinfo{person}{Rong Ge},
  {and} \bibinfo{person}{Xizhou Feng}.} \bibinfo{year}{2005}\natexlab{}.
\newblock \showarticletitle{High-performance, power-aware distributed computing
  for scientific applications}.
\newblock \bibinfo{journal}{\emph{Computer}} \bibinfo{volume}{38},
  \bibinfo{number}{11} (\bibinfo{year}{2005}), \bibinfo{pages}{40--47}.
\newblock


\bibitem[\protect\citeauthoryear{Castrillon, Leupers, and Ascheid}{Castrillon
  et~al\mbox{.}}{2011}]%
        {castrillon2011maps}
\bibfield{author}{\bibinfo{person}{Jeronimo Castrillon},
  \bibinfo{person}{Rainer Leupers}, {and} \bibinfo{person}{Gerd Ascheid}.}
  \bibinfo{year}{2011}\natexlab{}.
\newblock \showarticletitle{MAPS: Mapping concurrent dataflow applications to
  heterogeneous MPSoCs}.
\newblock \bibinfo{journal}{\emph{IEEE Transactions on Industrial Informatics}}
  \bibinfo{volume}{9}, \bibinfo{number}{1} (\bibinfo{year}{2011}),
  \bibinfo{pages}{527--545}.
\newblock


\bibitem[\protect\citeauthoryear{Desnos and Heulot}{Desnos and Heulot}{2014}]%
        {desnos2014pisdf}
\bibfield{author}{\bibinfo{person}{Karol Desnos} {and} \bibinfo{person}{Julien
  Heulot}.} \bibinfo{year}{2014}\natexlab{}.
\newblock \showarticletitle{Pisdf: Parameterized \& interfaced synchronous
  dataflow for mpsocs runtime reconfiguration}. In
  \bibinfo{booktitle}{\emph{1st Workshop on MEthods and TOols for Dataflow
  PrOgramming (METODO)}}.
\newblock


\bibitem[\protect\citeauthoryear{Desnos, Pelcat, Nezan, Bhattacharyya, and
  Aridhi}{Desnos et~al\mbox{.}}{2013}]%
        {desnos2013pimm}
\bibfield{author}{\bibinfo{person}{Karol Desnos}, \bibinfo{person}{Maxime
  Pelcat}, \bibinfo{person}{Jean-Fran{\c{c}}ois Nezan},
  \bibinfo{person}{Shuvra~S Bhattacharyya}, {and} \bibinfo{person}{Slaheddine
  Aridhi}.} \bibinfo{year}{2013}\natexlab{}.
\newblock \showarticletitle{Pimm: Parameterized and interfaced dataflow
  meta-model for mpsocs runtime reconfiguration}. In
  \bibinfo{booktitle}{\emph{2013 International Conference on Embedded Computer
  Systems: Architectures, Modeling, and Simulation (SAMOS)}}. IEEE,
  \bibinfo{pages}{41--48}.
\newblock


\bibitem[\protect\citeauthoryear{Desplanques, Demuynck, and
  Martens}{Desplanques et~al\mbox{.}}{2017}]%
        {desplanques2017adaptive}
\bibfield{author}{\bibinfo{person}{Brecht Desplanques}, \bibinfo{person}{Kris
  Demuynck}, {and} \bibinfo{person}{Jean-Pierre Martens}.}
  \bibinfo{year}{2017}\natexlab{}.
\newblock \showarticletitle{Adaptive speaker diarization of broadcast news
  based on factor analysis}.
\newblock \bibinfo{journal}{\emph{Computer Speech \& Language}}
  \bibinfo{volume}{46} (\bibinfo{year}{2017}), \bibinfo{pages}{72--93}.
\newblock


\bibitem[\protect\citeauthoryear{Eskandarian}{Eskandarian}{2012}]%
        {eskandarian2012handbook}
\bibfield{author}{\bibinfo{person}{Azim Eskandarian}.}
  \bibinfo{year}{2012}\natexlab{}.
\newblock \bibinfo{booktitle}{\emph{Handbook of intelligent vehicles}}.
  Vol.~\bibinfo{volume}{2}.
\newblock \bibinfo{publisher}{Springer}.
\newblock


\bibitem[\protect\citeauthoryear{Ge, Feng, Song, Chang, Li, and Cameron}{Ge
  et~al\mbox{.}}{2009}]%
        {ge2009powerpack}
\bibfield{author}{\bibinfo{person}{Rong Ge}, \bibinfo{person}{Xizhou Feng},
  \bibinfo{person}{Shuaiwen Song}, \bibinfo{person}{Hung-Ching Chang},
  \bibinfo{person}{Dong Li}, {and} \bibinfo{person}{Kirk~W Cameron}.}
  \bibinfo{year}{2009}\natexlab{}.
\newblock \showarticletitle{Powerpack: Energy profiling and analysis of
  high-performance systems and applications}.
\newblock \bibinfo{journal}{\emph{IEEE Transactions on Parallel and Distributed
  Systems}} \bibinfo{volume}{21}, \bibinfo{number}{5} (\bibinfo{year}{2009}),
  \bibinfo{pages}{658--671}.
\newblock


\bibitem[\protect\citeauthoryear{Ghahabi, Zhou, and Fischer}{Ghahabi
  et~al\mbox{.}}{2018}]%
        {ghahabi2018robust}
\bibfield{author}{\bibinfo{person}{Omid Ghahabi}, \bibinfo{person}{Wei Zhou},
  {and} \bibinfo{person}{Volker Fischer}.} \bibinfo{year}{2018}\natexlab{}.
\newblock \showarticletitle{A robust voice activity detection for real-time
  automatic speech recognition}.
\newblock \bibinfo{journal}{\emph{Studientexte zur Sprachkommunikation:
  Elektronische Sprachsignalverarbeitung 2018}} (\bibinfo{year}{2018}),
  \bibinfo{pages}{85--91}.
\newblock


\bibitem[\protect\citeauthoryear{Goens, Khasanov, Castrillon, H{\"a}hnel,
  Smejkal, and H{\"a}rtig}{Goens et~al\mbox{.}}{2017}]%
        {goens2017tetris}
\bibfield{author}{\bibinfo{person}{Andr{\'e}s Goens}, \bibinfo{person}{Robert
  Khasanov}, \bibinfo{person}{Jeronimo Castrillon}, \bibinfo{person}{Marcus
  H{\"a}hnel}, \bibinfo{person}{Till Smejkal}, {and} \bibinfo{person}{Hermann
  H{\"a}rtig}.} \bibinfo{year}{2017}\natexlab{}.
\newblock \showarticletitle{Tetris: A multi-application run-time system for
  predictable execution of static mappings}. In
  \bibinfo{booktitle}{\emph{Proceedings of SCOPES}}. ACM,
  \bibinfo{pages}{11--20}.
\newblock


\bibitem[\protect\citeauthoryear{Hellert, Koch, and St{\"u}tz}{Hellert
  et~al\mbox{.}}{2019}]%
        {hellert2019using}
\bibfield{author}{\bibinfo{person}{Christian Hellert}, \bibinfo{person}{Simon
  Koch}, {and} \bibinfo{person}{Peter St{\"u}tz}.}
  \bibinfo{year}{2019}\natexlab{}.
\newblock \showarticletitle{Using algorithm selection for adaptive vehicle
  perception aboard UAV}. In \bibinfo{booktitle}{\emph{2019 16th IEEE
  International Conference on Advanced Video and Signal Based Surveillance
  (AVSS)}}. IEEE, \bibinfo{pages}{1--8}.
\newblock


\bibitem[\protect\citeauthoryear{Istrate, Fredouille, Meignier, Besacier, and
  Bonastre}{Istrate et~al\mbox{.}}{2005}]%
        {istrate2005nist}
\bibfield{author}{\bibinfo{person}{Dan Istrate}, \bibinfo{person}{Corinne
  Fredouille}, \bibinfo{person}{Sylvain Meignier}, \bibinfo{person}{Laurent
  Besacier}, {and} \bibinfo{person}{Jean~Fran{\c{c}}ois Bonastre}.}
  \bibinfo{year}{2005}\natexlab{}.
\newblock \showarticletitle{NIST RT’05S evaluation: pre-processing techniques
  and speaker diarization on multiple microphone meetings}. In
  \bibinfo{booktitle}{\emph{International Workshop on Machine Learning for
  Multimodal Interaction}}. Springer, \bibinfo{pages}{428--439}.
\newblock


\bibitem[\protect\citeauthoryear{KAHN}{KAHN}{1974}]%
        {gilles1974semantics}
\bibfield{author}{\bibinfo{person}{Gilles KAHN}.}
  \bibinfo{year}{1974}\natexlab{}.
\newblock \showarticletitle{The semantics of a simple language for parallel
  programming}.
\newblock \bibinfo{journal}{\emph{In Information Processing}}
  \bibinfo{volume}{74} (\bibinfo{year}{1974}), \bibinfo{pages}{471--475}.
\newblock


\bibitem[\protect\citeauthoryear{Kashgarani and Kotthoff}{Kashgarani and
  Kotthoff}{2021}]%
        {kashgarani2021algorithm}
\bibfield{author}{\bibinfo{person}{Haniye Kashgarani} {and}
  \bibinfo{person}{Lars Kotthoff}.} \bibinfo{year}{2021}\natexlab{}.
\newblock \showarticletitle{Is algorithm selection worth it? Comparing
  selecting single algorithms and parallel execution}. In
  \bibinfo{booktitle}{\emph{AAAI Workshop on Meta-Learning and MetaDL
  Challenge}}. PMLR, \bibinfo{pages}{58--64}.
\newblock


\bibitem[\protect\citeauthoryear{Kerschke, Hoos, Neumann, and
  Trautmann}{Kerschke et~al\mbox{.}}{2019}]%
        {kerschke2019automated}
\bibfield{author}{\bibinfo{person}{Pascal Kerschke}, \bibinfo{person}{Holger~H
  Hoos}, \bibinfo{person}{Frank Neumann}, {and} \bibinfo{person}{Heike
  Trautmann}.} \bibinfo{year}{2019}\natexlab{}.
\newblock \showarticletitle{Automated algorithm selection: Survey and
  perspectives}.
\newblock \bibinfo{journal}{\emph{Evolutionary computation}}
  \bibinfo{volume}{27}, \bibinfo{number}{1} (\bibinfo{year}{2019}),
  \bibinfo{pages}{3--45}.
\newblock


\bibitem[\protect\citeauthoryear{Khasanov, Goens, and Castrillon}{Khasanov
  et~al\mbox{.}}{2018}]%
        {khasanov2018implicit}
\bibfield{author}{\bibinfo{person}{Robert Khasanov},
  \bibinfo{person}{Andr{\'e}s Goens}, {and} \bibinfo{person}{Jeronimo
  Castrillon}.} \bibinfo{year}{2018}\natexlab{}.
\newblock \showarticletitle{Implicit data-parallelism in {K}ahn process
  networks: Bridging the {M}ac{Q}ueen {G}ap}. In
  \bibinfo{booktitle}{\emph{Proceedings of PARMA-DITAM}}. ACM,
  \bibinfo{pages}{20--25}.
\newblock


\bibitem[\protect\citeauthoryear{Khasanov, Robledo, Menard, Goens, and
  Castrillon}{Khasanov et~al\mbox{.}}{2021}]%
        {khasanov_cases21}
\bibfield{author}{\bibinfo{person}{Robert Khasanov}, \bibinfo{person}{Julian
  Robledo}, \bibinfo{person}{Christian Menard}, \bibinfo{person}{Andres Goens},
  {and} \bibinfo{person}{Jeronimo Castrillon}.}
  \bibinfo{year}{2021}\natexlab{}.
\newblock \showarticletitle{Domain-specific hybrid mapping for energy-efficient
  baseband processing in wireless networks}.
\newblock \bibinfo{journal}{\emph{ACM Transactions on Embedded Computing
  Systems (TECS), special issue of the 2021 International Conference on
  Compilers, Architecture, and Synthesis of Embedded Systems (CASES)}}
  \bibinfo{volume}{20}, \bibinfo{number}{5s}, Article \bibinfo{articleno}{60}
  (\bibinfo{date}{Sept.} \bibinfo{year}{2021}), \bibinfo{numpages}{26}~pages.
\newblock
\showISSN{1539-9087}
\urldef\tempurl%
\url{https://doi.org/10.1145/3476991}
\showDOI{\tempurl}


\bibitem[\protect\citeauthoryear{Kotthoff, Gent, and Miguel}{Kotthoff
  et~al\mbox{.}}{2012}]%
        {kotthoff2012evaluation}
\bibfield{author}{\bibinfo{person}{Lars Kotthoff}, \bibinfo{person}{Ian~P
  Gent}, {and} \bibinfo{person}{Ian Miguel}.} \bibinfo{year}{2012}\natexlab{}.
\newblock \showarticletitle{An evaluation of machine learning in algorithm
  selection for search problems}.
\newblock \bibinfo{journal}{\emph{Ai Communications}} \bibinfo{volume}{25},
  \bibinfo{number}{3} (\bibinfo{year}{2012}), \bibinfo{pages}{257--270}.
\newblock


\bibitem[\protect\citeauthoryear{Kuutti, Fallah, Katsaros, Dianati, Mccullough,
  and Mouzakitis}{Kuutti et~al\mbox{.}}{2018}]%
        {kuutti2018survey}
\bibfield{author}{\bibinfo{person}{Sampo Kuutti}, \bibinfo{person}{Saber
  Fallah}, \bibinfo{person}{Konstantinos Katsaros}, \bibinfo{person}{Mehrdad
  Dianati}, \bibinfo{person}{Francis Mccullough}, {and}
  \bibinfo{person}{Alexandros Mouzakitis}.} \bibinfo{year}{2018}\natexlab{}.
\newblock \showarticletitle{A survey of the state-of-the-art localization
  techniques and their potentials for autonomous vehicle applications}.
\newblock \bibinfo{journal}{\emph{IEEE Internet of Things Journal}}
  \bibinfo{volume}{5}, \bibinfo{number}{2} (\bibinfo{year}{2018}),
  \bibinfo{pages}{829--846}.
\newblock


\bibitem[\protect\citeauthoryear{{Lee} and {Messerschmitt}}{{Lee} and
  {Messerschmitt}}{1987}]%
        {lee87staticScedDfDsp}
\bibfield{author}{\bibinfo{person}{E.~A. {Lee}} {and} \bibinfo{person}{D.~G.
  {Messerschmitt}}.} \bibinfo{year}{1987}\natexlab{}.
\newblock \showarticletitle{Static Scheduling of Synchronous Data Flow Programs
  for Digital Signal Processing}.
\newblock \bibinfo{journal}{\emph{IEEE Trans. Comput.}} \bibinfo{volume}{C-36},
  \bibinfo{number}{1} (\bibinfo{year}{1987}), \bibinfo{pages}{24--35}.
\newblock


\bibitem[\protect\citeauthoryear{Lee and Messerschmitt}{Lee and
  Messerschmitt}{1987}]%
        {lee87sdf}
\bibfield{author}{\bibinfo{person}{Edward~A. Lee} {and}
  \bibinfo{person}{David~G. Messerschmitt}.} \bibinfo{year}{1987}\natexlab{}.
\newblock \showarticletitle{{Synchronous Data Flow}}.
\newblock \bibinfo{journal}{\emph{Proc. IEEE}} \bibinfo{volume}{75},
  \bibinfo{number}{9} (\bibinfo{date}{Sept.} \bibinfo{year}{1987}),
  \bibinfo{pages}{1235--1245}.
\newblock
\showISSN{0018-9219}


\bibitem[\protect\citeauthoryear{{Lee} and {Messerschmitt}}{{Lee} and
  {Messerschmitt}}{1987}]%
        {1458143}
\bibfield{author}{\bibinfo{person}{E.~A. {Lee}} {and} \bibinfo{person}{D.~G.
  {Messerschmitt}}.} \bibinfo{year}{1987}\natexlab{}.
\newblock \showarticletitle{Synchronous data flow}.
\newblock \bibinfo{journal}{\emph{Proc. IEEE}} \bibinfo{volume}{75},
  \bibinfo{number}{9} (\bibinfo{date}{Sept.} \bibinfo{year}{1987}),
  \bibinfo{pages}{1235--1245}.
\newblock


\bibitem[\protect\citeauthoryear{Lee and Sangiovanni-Vincentelli}{Lee and
  Sangiovanni-Vincentelli}{1998}]%
        {lee1998framework}
\bibfield{author}{\bibinfo{person}{Edward~A. Lee} {and}
  \bibinfo{person}{Alberto Sangiovanni-Vincentelli}.}
  \bibinfo{year}{1998}\natexlab{}.
\newblock \showarticletitle{A framework for comparing models of computation}.
\newblock \bibinfo{journal}{\emph{IEEE Transactions on computer-aided design of
  integrated circuits and systems}} \bibinfo{volume}{17}, \bibinfo{number}{12}
  (\bibinfo{year}{1998}), \bibinfo{pages}{1217--1229}.
\newblock


\bibitem[\protect\citeauthoryear{Lv, Guan, Zhang, Deng, Yu, and Zhang}{Lv
  et~al\mbox{.}}{2009}]%
        {lv2009survey}
\bibfield{author}{\bibinfo{person}{Mingsong Lv}, \bibinfo{person}{Nan Guan},
  \bibinfo{person}{Yi Zhang}, \bibinfo{person}{Qingxu Deng},
  \bibinfo{person}{Ge Yu}, {and} \bibinfo{person}{Jianming Zhang}.}
  \bibinfo{year}{2009}\natexlab{}.
\newblock \showarticletitle{A survey of WCET analysis of real-time operating
  systems}. In \bibinfo{booktitle}{\emph{2009 International Conference on
  Embedded Software and Systems}}. IEEE, \bibinfo{pages}{65--72}.
\newblock


\bibitem[\protect\citeauthoryear{Madronal, Arrestier, Sancho, Morvan, Lazcano,
  Desnos, Salvador, Menard, Juarez, and Sanz}{Madronal et~al\mbox{.}}{2019}]%
        {madronal2019papify}
\bibfield{author}{\bibinfo{person}{Daniel Madronal}, \bibinfo{person}{Florian
  Arrestier}, \bibinfo{person}{Jaime Sancho}, \bibinfo{person}{Antoine Morvan},
  \bibinfo{person}{Raquel Lazcano}, \bibinfo{person}{Karol Desnos},
  \bibinfo{person}{Ruben Salvador}, \bibinfo{person}{Daniel Menard},
  \bibinfo{person}{Eduardo Juarez}, {and} \bibinfo{person}{Cesar Sanz}.}
  \bibinfo{year}{2019}\natexlab{}.
\newblock \showarticletitle{Papify: Automatic instrumentation and monitoring of
  dynamic dataflow applications based on papi}.
\newblock \bibinfo{journal}{\emph{IEEE Access}}  \bibinfo{volume}{7}
  (\bibinfo{year}{2019}), \bibinfo{pages}{111801--111812}.
\newblock


\bibitem[\protect\citeauthoryear{Meduri and Ananth}{Meduri and Ananth}{2012}]%
        {meduri2012survey}
\bibfield{author}{\bibinfo{person}{Seshashyama~Sameeraj Meduri} {and}
  \bibinfo{person}{Rufus Ananth}.} \bibinfo{year}{2012}\natexlab{}.
\newblock \bibinfo{title}{A survey and evaluation of voice activity detection
  algorithms}.
\newblock
\newblock


\bibitem[\protect\citeauthoryear{Mohamed, Haghbayan, Westerlund, Heikkonen,
  Tenhunen, and Plosila}{Mohamed et~al\mbox{.}}{2019}]%
        {mohamed2019survey}
\bibfield{author}{\bibinfo{person}{Sherif~AS Mohamed},
  \bibinfo{person}{Mohammad-Hashem Haghbayan}, \bibinfo{person}{Tomi
  Westerlund}, \bibinfo{person}{Jukka Heikkonen}, \bibinfo{person}{Hannu
  Tenhunen}, {and} \bibinfo{person}{Juha Plosila}.}
  \bibinfo{year}{2019}\natexlab{}.
\newblock \showarticletitle{A survey on odometry for autonomous navigation
  systems}.
\newblock \bibinfo{journal}{\emph{IEEE Access}}  \bibinfo{volume}{7}
  (\bibinfo{year}{2019}), \bibinfo{pages}{97466--97486}.
\newblock


\bibitem[\protect\citeauthoryear{Munoz, Kirley, and Halgamuge}{Munoz
  et~al\mbox{.}}{2013}]%
        {munoz2013algorithm}
\bibfield{author}{\bibinfo{person}{Mario~A Munoz}, \bibinfo{person}{Michael
  Kirley}, {and} \bibinfo{person}{Saman~K Halgamuge}.}
  \bibinfo{year}{2013}\natexlab{}.
\newblock \showarticletitle{The algorithm selection problem on the continuous
  optimization domain}.
\newblock In \bibinfo{booktitle}{\emph{Computational intelligence in
  intelligent data analysis}}. \bibinfo{publisher}{Springer},
  \bibinfo{pages}{75--89}.
\newblock


\bibitem[\protect\citeauthoryear{Nosratighods, Ambikairajah, and
  Epps}{Nosratighods et~al\mbox{.}}{2006}]%
        {nosratighods2006speaker}
\bibfield{author}{\bibinfo{person}{Mohaddeseh Nosratighods},
  \bibinfo{person}{Eliathamby Ambikairajah}, {and} \bibinfo{person}{Julien
  Epps}.} \bibinfo{year}{2006}\natexlab{}.
\newblock \showarticletitle{Speaker verification using a novel set of dynamic
  features}. In \bibinfo{booktitle}{\emph{18th International Conference on
  Pattern Recognition (ICPR'06)}}, Vol.~\bibinfo{volume}{4}. IEEE,
  \bibinfo{pages}{266--269}.
\newblock


\bibitem[\protect\citeauthoryear{PreesmProject}{PreesmProject}{2020}]%
        {Preesm}
\bibfield{author}{\bibinfo{person}{PreesmProject}.} \bibinfo{year}{(last
  accessed November 11, 2020)}\natexlab{}.
\newblock \bibinfo{booktitle}{\emph{\href{https://preesm.github.io/}{Preesm}}}.
\newblock
\urldef\tempurl%
\url{https://preesm.github.io/}
\showURL{%
\tempurl}


\bibitem[\protect\citeauthoryear{Ptolemaeus}{Ptolemaeus}{2014}]%
        {ptolemyBook}
\bibfield{author}{\bibinfo{person}{Claudius Ptolemaeus}.}
  \bibinfo{year}{2014}\natexlab{}.
\newblock \bibinfo{booktitle}{\emph{System Design, Modeling, and Simulation
  using Ptolemy II, 2014}}.
\newblock \bibinfo{publisher}{Ptolemy.org}.
\newblock


\bibitem[\protect\citeauthoryear{Rice et~al\mbox{.}}{Rice
  et~al\mbox{.}}{1976}]%
        {rice1976algorithm}
\bibfield{author}{\bibinfo{person}{John~R. Rice} {et~al\mbox{.}}}
  \bibinfo{year}{1976}\natexlab{}.
\newblock \showarticletitle{The algorithm selection problem}.
\newblock \bibinfo{journal}{\emph{Advances in computers}} \bibinfo{volume}{15},
  \bibinfo{number}{65-118} (\bibinfo{year}{1976}), \bibinfo{pages}{5}.
\newblock


\bibitem[\protect\citeauthoryear{Rubattu, Palumbo, Bhattacharyya, and
  Pelcat}{Rubattu et~al\mbox{.}}{2022}]%
        {rubattu2022pathtracer}
\bibfield{author}{\bibinfo{person}{Claudio Rubattu}, \bibinfo{person}{Francesca
  Palumbo}, \bibinfo{person}{Shuvra~S Bhattacharyya}, {and}
  \bibinfo{person}{Maxime Pelcat}.} \bibinfo{year}{2022}\natexlab{}.
\newblock \showarticletitle{PathTracer: Understanding Response Time of Signal
  Processing Applications on Heterogeneous MPSoCs}.
\newblock \bibinfo{journal}{\emph{ACM Transactions on Modeling and Performance
  Evaluation of Computing Systems}} (\bibinfo{year}{2022}).
\newblock


\bibitem[\protect\citeauthoryear{Ryant, Church, Cieri, Cristia, Du, Ganapathy,
  and Liberman}{Ryant et~al\mbox{.}}{2018}]%
        {ryant2018first}
\bibfield{author}{\bibinfo{person}{Neville Ryant}, \bibinfo{person}{Kenneth
  Church}, \bibinfo{person}{Christopher Cieri}, \bibinfo{person}{Alejandrina
  Cristia}, \bibinfo{person}{Jun Du}, \bibinfo{person}{Sriram Ganapathy}, {and}
  \bibinfo{person}{Mark Liberman}.} \bibinfo{year}{2018}\natexlab{}.
\newblock \showarticletitle{First DIHARD challenge evaluation plan}.
\newblock \bibinfo{journal}{\emph{2018, tech. Rep.}} (\bibinfo{year}{2018}).
\newblock


\bibitem[\protect\citeauthoryear{Sahayadhas, Sundaraj, and
  Murugappan}{Sahayadhas et~al\mbox{.}}{2012}]%
        {sahayadhas2012detecting}
\bibfield{author}{\bibinfo{person}{Arun Sahayadhas}, \bibinfo{person}{Kenneth
  Sundaraj}, {and} \bibinfo{person}{Murugappan Murugappan}.}
  \bibinfo{year}{2012}\natexlab{}.
\newblock \showarticletitle{Detecting driver drowsiness based on sensors: a
  review}.
\newblock \bibinfo{journal}{\emph{Sensors}} \bibinfo{volume}{12},
  \bibinfo{number}{12} (\bibinfo{year}{2012}), \bibinfo{pages}{16937--16953}.
\newblock


\bibitem[\protect\citeauthoryear{Schor, Bacivarov, Yang, and Thiele}{Schor
  et~al\mbox{.}}{2014a}]%
        {schor2014adapnet}
\bibfield{author}{\bibinfo{person}{Lars Schor}, \bibinfo{person}{Iuliana
  Bacivarov}, \bibinfo{person}{Hoeseok Yang}, {and} \bibinfo{person}{Lothar
  Thiele}.} \bibinfo{year}{2014}\natexlab{a}.
\newblock \showarticletitle{Adapnet: Adapting process networks in response to
  resource variations}. In \bibinfo{booktitle}{\emph{Proceedings of CASES}}.
  ACM, \bibinfo{pages}{22}.
\newblock


\bibitem[\protect\citeauthoryear{Schor, Bacivarov, Yang, and Thiele}{Schor
  et~al\mbox{.}}{2014b}]%
        {Schor2014Oct12}
\bibfield{author}{\bibinfo{person}{Lars Schor}, \bibinfo{person}{Iuliana
  Bacivarov}, \bibinfo{person}{Hoeseok Yang}, {and} \bibinfo{person}{Lothar
  Thiele}.} \bibinfo{year}{2014}\natexlab{b}.
\newblock \showarticletitle{Ada{PN}et: {A}dapting process networks in response
  to resource variations}. In \bibinfo{booktitle}{\emph{Proceedings of the 2014
  International Conference on Compilers, Architecture and Synthesis for
  Embedded Systems}}. \bibinfo{pages}{1--10}.
\newblock


\bibitem[\protect\citeauthoryear{Singh, Shafique, Kumar, and Henkel}{Singh
  et~al\mbox{.}}{2013}]%
        {2013dacSingh}
\bibfield{author}{\bibinfo{person}{A.K. Singh}, \bibinfo{person}{M. Shafique},
  \bibinfo{person}{A. Kumar}, {and} \bibinfo{person}{J. Henkel}.}
  \bibinfo{year}{2013}\natexlab{}.
\newblock \showarticletitle{Mapping on multi/many-core systems:Survey of
  current and emerging trends}. In \bibinfo{booktitle}{\emph{Proceedings of
  50th ACM/EDAC/IEEE DAC'13}}. \bibinfo{pages}{1--10}.
\newblock
\showISSN{0738-100X}


\bibitem[\protect\citeauthoryear{Stuijk, Geilen, and Basten}{Stuijk
  et~al\mbox{.}}{2006}]%
        {stuijk2006sdf}
\bibfield{author}{\bibinfo{person}{Sander Stuijk}, \bibinfo{person}{Marc
  Geilen}, {and} \bibinfo{person}{Twan Basten}.}
  \bibinfo{year}{2006}\natexlab{}.
\newblock \showarticletitle{Sdf3: {SDF} for free}. In
  \bibinfo{booktitle}{\emph{Sixth International Conference on Application of
  Concurrency to System Design (ACSD'06)}}. IEEE, \bibinfo{pages}{276--278}.
\newblock


\bibitem[\protect\citeauthoryear{Stuijk, Geilen, Theelen, and Basten}{Stuijk
  et~al\mbox{.}}{2011}]%
        {stuijk2011scenario}
\bibfield{author}{\bibinfo{person}{Sander Stuijk}, \bibinfo{person}{Marc
  Geilen}, \bibinfo{person}{Bart Theelen}, {and} \bibinfo{person}{Twan
  Basten}.} \bibinfo{year}{2011}\natexlab{}.
\newblock \showarticletitle{Scenario-aware dataflow: Modeling, analysis and
  implementation of dynamic applications}. In
  \bibinfo{booktitle}{\emph{Proceedings of SAMOS}}. IEEE,
  \bibinfo{pages}{404--411}.
\newblock


\bibitem[\protect\citeauthoryear{Togneri and Pullella}{Togneri and
  Pullella}{2011}]%
        {togneri2011overview}
\bibfield{author}{\bibinfo{person}{Roberto Togneri} {and}
  \bibinfo{person}{Daniel Pullella}.} \bibinfo{year}{2011}\natexlab{}.
\newblock \showarticletitle{An overview of speaker identification: Accuracy and
  robustness issues}.
\newblock \bibinfo{journal}{\emph{IEEE Circuits and Systems Magazine}}
  \bibinfo{volume}{11}, \bibinfo{number}{2} (\bibinfo{year}{2011}),
  \bibinfo{pages}{23--61}.
\newblock


\bibitem[\protect\citeauthoryear{Tuzun, Demirekler, and Nakiboglu}{Tuzun
  et~al\mbox{.}}{1994}]%
        {tuzun1994comparison}
\bibfield{author}{\bibinfo{person}{OB Tuzun}, \bibinfo{person}{M Demirekler},
  {and} \bibinfo{person}{KB Nakiboglu}.} \bibinfo{year}{1994}\natexlab{}.
\newblock \showarticletitle{Comparison of parametric and non-parametric
  representations of speech for recognition}. In
  \bibinfo{booktitle}{\emph{Electrotechnical Conference, 1994. Proceedings.,
  7th Mediterranean}}. IEEE, \bibinfo{pages}{65--68}.
\newblock


\bibitem[\protect\citeauthoryear{Wang, Du, He, Niu, Sun, and Lee}{Wang
  et~al\mbox{.}}{2021}]%
        {wang2021scenario}
\bibfield{author}{\bibinfo{person}{Yu-Xuan Wang}, \bibinfo{person}{Jun Du},
  \bibinfo{person}{Maokui He}, \bibinfo{person}{Shu-Tong Niu},
  \bibinfo{person}{Lei Sun}, {and} \bibinfo{person}{Chin-Hui Lee}.}
  \bibinfo{year}{2021}\natexlab{}.
\newblock \showarticletitle{Scenario-dependent speaker diarization for
  dihard-iii challenge}.
\newblock \bibinfo{journal}{\emph{Accepted to Interspeech}}
  (\bibinfo{year}{2021}).
\newblock


\bibitem[\protect\citeauthoryear{Wolpert and Macready}{Wolpert and
  Macready}{1997}]%
        {wolpert1997no}
\bibfield{author}{\bibinfo{person}{David~H Wolpert} {and}
  \bibinfo{person}{William~G Macready}.} \bibinfo{year}{1997}\natexlab{}.
\newblock \showarticletitle{No free lunch theorems for optimization}.
\newblock \bibinfo{journal}{\emph{IEEE transactions on evolutionary
  computation}} \bibinfo{volume}{1}, \bibinfo{number}{1}
  (\bibinfo{year}{1997}), \bibinfo{pages}{67--82}.
\newblock


\bibitem[\protect\citeauthoryear{Yin}{Yin}{2019}]%
        {yin2019steps}
\bibfield{author}{\bibinfo{person}{Ruiqing Yin}.}
  \bibinfo{year}{2019}\natexlab{}.
\newblock \emph{\bibinfo{title}{Steps towards end-to-end neural speaker
  diarization}}.
\newblock \bibinfo{thesistype}{Ph.D. Dissertation}. \bibinfo{school}{Paris
  Saclay}.
\newblock


\bibitem[\protect\citeauthoryear{Yu and Deng}{Yu and Deng}{2016}]%
        {yu2016automatic}
\bibfield{author}{\bibinfo{person}{Dong Yu} {and} \bibinfo{person}{Li Deng}.}
  \bibinfo{year}{2016}\natexlab{}.
\newblock \bibinfo{booktitle}{\emph{AUTOMATIC SPEECH RECOGNITION.}}
\newblock \bibinfo{publisher}{Springer}.
\newblock


\bibitem[\protect\citeauthoryear{Yurtsever, Lambert, Carballo, and
  Takeda}{Yurtsever et~al\mbox{.}}{2020}]%
        {yurtsever2020survey}
\bibfield{author}{\bibinfo{person}{Ekim Yurtsever}, \bibinfo{person}{Jacob
  Lambert}, \bibinfo{person}{Alexander Carballo}, {and} \bibinfo{person}{Kazuya
  Takeda}.} \bibinfo{year}{2020}\natexlab{}.
\newblock \showarticletitle{A survey of autonomous driving: Common practices
  and emerging technologies}.
\newblock \bibinfo{journal}{\emph{IEEE access}}  \bibinfo{volume}{8}
  (\bibinfo{year}{2020}), \bibinfo{pages}{58443--58469}.
\newblock


\end{thebibliography}

\end{document}